\documentclass[letterpaper,onecolumn]{IEEEtran}
\usepackage{amsmath,amsfonts}
\usepackage{algorithm}
\usepackage{algpseudocode}
\usepackage{array}
\usepackage[caption=false,font=normalsize,labelfont=sf,textfont=sf]{subfig}
\usepackage{multirow}
\usepackage{textcomp}
\usepackage{stfloats}
\usepackage{url}
\usepackage{verbatim}
\usepackage{graphicx}
\usepackage{cite}
\usepackage{amsthm}
\usepackage{booktabs}
\usepackage{enumitem}
\usepackage{threeparttable}
\usepackage{balance}
\usepackage{hyperref}

\newcommand{\multirowfit}{*}

\begin{document}

\title{Scalable and Private Federated Learning Using Distributed Differential Privacy and Secure Aggregation}

\author{Wenjing~Wei,
        Farid~Nait-Abdesselam
        and~Alla~Jammine%
\thanks{W. Wei, F. Nait-Abdesselam and A. Jammine are with Université Paris Cité, Paris, France (e-mail: wenjing.wei@etu.u-paris.fr, farid.nait-abdesselam@u-paris.fr, and alla.jammine@u-paris.fr).}%
}

\markboth{}%
{Wei \MakeLowercase{\textit{et al.}}: Scalable and Private Federated Learning Using Distributed Differential Privacy and Secure Aggregation}

\IEEEpubid{0000--0000/00\$00.00~\copyright~2025 IEEE}

\maketitle


\begin{abstract}
This article presents \textsc{DDP-SA}, a scalable privacy-preserving federated learning framework that jointly leverages client-side local differential privacy (LDP) and full-threshold additive secret sharing (ASS) for secure aggregation. Unlike existing methods that rely solely on differential privacy or on secure multi-party computation (MPC), \textsc{DDP-SA} integrates both techniques to deliver stronger end-to-end privacy guarantees while remaining computationally practical. The framework introduces a two-stage protection mechanism: clients first perturb their local gradients with calibrated Laplace noise, then decompose the noisy gradients into additive secret shares that are distributed across multiple intermediate servers. This design ensures that (i) no single compromised server or communication channel can reveal any information about individual client updates, and (ii) the parameter server reconstructs only the aggregated noisy gradient, never any client-specific contribution. Extensive experiments show that \textsc{DDP-SA} achieves substantially higher model accuracy than standalone LDP while providing stronger privacy protection than MPC-only approaches. The proposed framework scales linearly with the number of participants and offers a practical, privacy-preserving solution for federated learning applications with controllable computational and communication overhead.
\end{abstract}

\begin{IEEEkeywords}
Federated Learning, Differential Privacy, Secure Aggregation, Secret Sharing, Privacy-Preserving Machine Learning
\end{IEEEkeywords}

\section{Introduction}\label{sec:section1}

\IEEEPARstart{M}{achine} learning (ML) plays a central role in modern society and is widely adopted across numerous industries. It underpins applications in computer vision, speech recognition, natural language processing, and many other domains that significantly benefit users and organizations. Traditionally, ML systems require raw data to be uploaded from users' devices to a central server for model training. However, this centralized paradigm raises substantial privacy concerns, as it exposes sensitive user information to potential leakage~\cite{yang2019federated}.

To address these privacy and security challenges, \emph{federated learning} (FL) has emerged as a promising distributed ML framework~\cite{mcmahan2017communication}. FL enables multiple clients (e.g., mobile devices) to collaboratively train a global model by transmitting only locally computed updates, such as gradients or model parameters, while keeping raw data on-device. Although this paradigm provides an initial layer of privacy protection, recent studies have demonstrated that FL remains vulnerable to privacy leakage, particularly through inference attacks that exploit shared updates~\cite{phong2017privacy,nasr2019comprehensive}.

\IEEEpubidadjcol

These privacy risks predominantly arise from \emph{privacy inference attacks}, in which adversaries analyze shared updates to infer sensitive attributes of users’ data~\cite{kairouz2021advances,lee2021defensive}. Existing defenses, most notably differential privacy (DP) and secure multi-party computation (MPC), provide partial mitigation but exhibit notable limitations~\cite{dwork2014algorithmic}. Differential privacy obscures client updates by adding randomized noise, but stronger privacy requires larger noise magnitudes that significantly degrade model performance. In contrast, MPC-based secure aggregation protocols cryptographically ensure that the server learns only aggregated results~\cite{bonawitz2017practical,xu2019verifynet,aono2017privacy,hao2019towards,hao2019efficient}, yet they often incur substantial computational and communication overhead.

Motivated by these challenges and the limitations of using DP or MPC alone, we propose a novel privacy-preserving federated learning framework, \emph{Distributed Differential Privacy via Secure Aggregation} (DDP-SA). DDP-SA integrates client-side local differential privacy (LDP) with full-threshold additive secret sharing (ASS), resulting in a principled hybrid mechanism that achieves formal $(\epsilon,\delta)$ differential privacy at the client level while cryptographically hiding individual updates from both the server and all communication paths. As established in Theorem~\ref{ddp-sa-privacy}, the combined mechanism retains its DP guarantee due to post-processing invariance while ensuring that no single client’s contribution is ever exposed.

To support scalability, we design a multi-server architecture consisting of $n$ clients and $m$ intermediate servers. This architecture achieves linear communication complexity and generalizes naturally to arbitrary $m$, extending beyond commonly studied illustrative cases such as $m=3$. Within this architecture, clients first perturb their gradients with calibrated Laplace noise, then encode the noisy gradients into additive secret shares that are distributed among the intermediate servers. These servers aggregate the received shares and forward only the combined result to the parameter server (PS), which reconstructs the aggregated noisy gradient and updates the global model.

Another key contribution of our work is a multi-round privacy analysis based on advanced composition. We provide practical guidance for allocating privacy budgets in long-running FL scenarios, which enables system designers to manage privacy loss across many training rounds. Additionally, we conduct a detailed component-wise breakdown of computational and communication costs, separating the overhead introduced by the LDP and MPC components. Our analysis highlights the specific sources of system overhead and demonstrates that secure aggregation remains practical even at a large scale.
In the entire FL process, clients never reveal raw data or unprotected gradients, providing resilience against a broad class of privacy inference attacks. Experimental results show that DDP-SA offers stronger privacy guarantees than either LDP or MPC alone, while maintaining acceptable accuracy and efficiency. Moreover, DDP-SA scales effectively to large numbers of clients and servers.

The structure of this paper is organized as follows. Section~\ref{sec:section1} introduces the research background, motivation, and main contributions. Section~\ref{sec:section2} reviews related work. Section~\ref{sec:section3} presents preliminaries. Section~\ref{sec:section4} describes the system overview and details of the DDP-SA framework. Section~\ref{sec:section5} provides privacy analysis. Section~\ref{sec:section6} presents experimental results and performance evaluation. Section~\ref{sec:section7} concludes the paper.

\section{Related Work}\label{sec:section2}

In recent years, FL has emerged as a powerful distributed machine learning paradigm that allows multiple participants to collaboratively train a global model without directly sharing their raw data. While FL offers promising privacy benefits compared to traditional centralized training, it remains vulnerable to a range of privacy inference attacks that can compromise sensitive client information. To address these risks, a growing body of research has focused on integrating advanced privacy-preserving techniques into FL systems, including differential privacy, secure multi-party computation, and homomorphic encryption (HE).

This section provides a comprehensive overview of the current landscape in privacy-preserving federated learning. We begin in Section~\ref{sec:section2.1} by categorizing various privacy inference attacks that threaten FL systems and highlighting their mechanisms and impact. We then explore the application of differential privacy in FL and examine its practical implementations and limitations. In addition, we discuss the role of secure computation techniques, particularly MPC and homomorphic encryption, in safeguarding model updates. The section further reviews recent hybrid approaches that combine DP and MPC to balance privacy, efficiency, and model accuracy.

Through this survey of the state of the art, we aim to contextualize the design and motivation behind our proposed privacy-preserving FL framework introduced in the subsequent sections.

\subsection{Privacy Inference Attacks in FL}\label{sec:section2.1}
Federated learning, as a distributed machine learning paradigm, can effectively address the privacy challenges faced by traditional centralized machine learning and has been widely adopted in areas involving users' sensitive data, such as healthcare, finance, and the Internet of Things (IoT). The federated averaging (FedAvg) algorithm is the core algorithm of FL. It includes both the model averaging algorithm and the gradient averaging algorithm~\cite{mcmahan2017communication,McMahan2016FederatedLO}. In the model averaging algorithm, users train their local models using stochastic gradient descent (SGD) and send the model parameters to the parameter server (PS) for aggregation. In the gradient averaging algorithm, users upload their gradient parameters to the PS for aggregation. The model averaging algorithm typically requires fewer communication rounds to reach convergence compared to the gradient averaging algorithm.

Even though FL provides some privacy protection compared to traditional machine learning, recent studies~\cite{nasr2019comprehensive,kairouz2021advances,lee2021defensive,melis2019exploiting,fredrikson2015model,zhu2019deep,zhao2020idlg,hitaj2017deep,geiping2020inverting} have shown that attackers can still obtain private information about users by analyzing exchanged model parameters or gradients. Melis et al.~\cite{melis2019exploiting} revealed an attack strategy that exploits unintended feature leakage from gradients shared during collaborative learning, allowing adversaries to infer sensitive attributes about participants' data without direct access to it. Fredrikson et al.~\cite{fredrikson2015model} presented model inversion attacks that use confidence information revealed by machine learning models to reconstruct sensitive input data, highlighting the privacy risks associated with exposing high-confidence predictions. In~\cite{zhu2019deep}, the authors demonstrated an attack in which adversaries can recover original training data from shared model gradients during the training process in deep learning, underscoring the significant privacy risks of gradient sharing in collaborative learning environments. Hitaj et al.~\cite{hitaj2017deep} showed that adversaries can use generative adversarial networks (GANs) to reconstruct private training data of other participants by exploiting shared model updates in collaborative deep learning settings.

\subsection{Differential Privacy in FL}\label{sec:section2.2}
With increased research interest in differential privacy, many researchers have applied various forms of DP, including central differential privacy, local differential privacy, and distributed differential privacy, to the federated learning process to defend against privacy inference attacks~\cite{hu2020personalized,geyer2017differentially,wei2020federated,liu2020fedsel,zhao2020local,zheng2021federated,seif2020wireless,chen2024differentially,malekmohammadi2024noise,liu2024cross,ling2024ali,ruan2023private,noble2022differentially}. Table~\ref{tab: An overview study of DP-FL} summarizes over 70 recent articles on differentially private FL. Hu et al.~\cite{hu2020personalized} introduced personalized federated learning with differential privacy, combining personalized model training with DP to improve data privacy and model performance. However, this approach may increase computational complexity and reduce model accuracy due to the noise added for privacy preservation. Geyer et al.~\cite{geyer2017differentially} proposed a client-level differentially private federated learning method that integrates DP directly into the FL process, although the added noise can degrade learning performance and reduce accuracy. Wei et al.~\cite{wei2020federated} developed federated learning algorithms incorporating differential privacy to safeguard user data privacy while enabling collaborative model training. A limitation of these algorithms is the inherent privacy-accuracy trade-off, since higher privacy levels typically reduce model accuracy.

Liu et al.~\cite{liu2020fedsel} introduced FedSel, a method combining federated SGD with local differential privacy and top-$k$ dimension selection, improving data privacy and training efficiency. However, selecting the top-$k$ dimensions may lead to information loss and reduced accuracy. Zhao et al.~\cite{zhao2020local} applied local differential privacy to protect IoT device data in FL while collectively improving model learning, although higher privacy levels can significantly impact learning effectiveness and convergence. Zheng et al.~\cite{zheng2021federated} introduced federated $f$-differential privacy, a flexible DP framework tailored for FL, but its implementation requires careful and sometimes complex privacy parameter selection. Seif et al.~\cite{seif2020wireless} proposed wireless federated learning combined with local differential privacy for secure user data protection in distributed training over wireless networks. However, increased noise and unreliable wireless transmission can reduce the accuracy of the federated model. All of the above DP-based schemes share a common limitation, since adding random noise to gradients or parameters inevitably decreases the accuracy of the federated learning model.

\begin{table*}[t!]
\caption{An Overview Study of Differentially Private FL~\cite{fu2024differentially}}
\label{tab: An overview study of DP-FL}
\tiny
\resizebox{\textwidth}{!}{
\centering
\begin{tabular}{|c|c|c|c|c|c|c|c|c|c|c|c|}
\hline
\textbf{\begin{tabular}[c]{@{}c@{}}Federated\\  Scenario\end{tabular}} & \textbf{Publications}                                & \textbf{Year} & \textbf{\begin{tabular}[c]{@{}c@{}}DP \\ Model\end{tabular}} & \textbf{\begin{tabular}[c]{@{}c@{}}Neighborhood\\ Level\end{tabular}} & \textbf{\begin{tabular}[c]{@{}c@{}}Perturbation \\ Mechanism\end{tabular}} & $\textbf{CM}^1$ & \textbf{\begin{tabular}[c]{@{}c@{}}Downstream\\  Tasks\end{tabular}} & \textbf{$\begin{tabular}[c]{@{}c@{}}Model \\ Architecture\end{tabular}^2$} & \textbf{\begin{tabular}[c]{@{}c@{}}Clients\\  Number\end{tabular}} & \textbf{$\epsilon$} & \textbf{$\delta$}                  \\ \hline
                                                                       & Chen et al.\cite{chen2024differentially}             & 2024          &                                                              &                                                                       & Gaussian                                                                   & tCDP                                                                     & Classification                                                      & LR, Shallow CNN                                                       & 100                                                                & 0.3                 & $10^{-2}$                          \\ \cline{2-3} \cline{6-12} 
                                                                       & malekmohammadi et al. \cite{malekmohammadi2024noise} & 2024          &                                                              &                                                                       & Gaussian                                                                   & AC                                                                       & Classification                                                      & CNN                                                                   & [20,60]                                                            & [0.5,5]             & $10^{-4}$                          \\ \cline{2-3} \cline{6-12} 
                                                                       & Liu et al.\cite{liu2024cross}                        & 2024          &                                                              &                                                                       & Gaussian                                                                   & RDP                                                                      & Classification                                                      & CNN                                                                   & 10                                                                 & [0.1,10]            & $10^{-3}$                          \\ \cline{2-3} \cline{6-12} 
                                                                       & Ling et al. \cite{ling2024ali}                       & 2024          &                                                              &                                                                       & Gaussian                                                                   & RDP                                                                      & Classification                                                      & Shallow CNN                                                           & 10                                                                 & [1.5,5.5]           & $10^{-5}$                          \\ \cline{2-3} \cline{6-12} 
                                                                       & Xiang et al. \cite{xiang2023practical}               & 2023          &                                                              &                                                                       & Gaussian                                                                   & MA                                                                       & Classification                                                      & Shallow CNN,LSTM                                                      & [10,20]                                                            & [0.12,2]            & $[10^{-2},10^{-5}]$                \\ \cline{2-3} \cline{6-12} 
                                                                       & Ruan et al. \cite{ruan2023private}                   & 2023          &                                                              &                                                                       & Gaussian                                                                   & RDP                                                                      & Classification                                                      & Shallow CNN, LSTM                                                     & [3,10]                                                             & [0.25,2]            & $[10^{-4},10^{-5}]$                \\ \cline{2-3} \cline{6-12} 
                                                                       & Noble et al. \cite{noble2022differentially}          & 2022          &                                                              &                                                                       & Gaussian                                                                   & RDP                                                                      & Classification                                                      & Shallow CNN                                                           & 10                                                                 & [3,13]              & $10^{-6}$                          \\ \cline{2-3} \cline{6-12} 
                                                                       & Fu et al. \cite{fu2022adap}                          & 2022          &                                                              &                                                                       & Gaussian                                                                   & RDP                                                                      & Classification                                                      & Shallow CNN                                                           & 10                                                                 & [2,6]               & $10^{-5}$                          \\ \cline{2-3} \cline{6-12} 
                                                                       & Li et al. \cite{li2022soteriafl}                     & 2022          &                                                              &                                                                       & Gaussian                                                                   & MA                                                                       & Classification                                                      & LR, Shallow CNN                                                       & 10                                                                 & [1,16]              & $10^{-3}$                          \\ \cline{2-3} \cline{6-12} 
                                                                       & Ryu et al. \cite{ryu2022differentially}              & 2022          &                                                              &                                                                       & Gaussian                                                                   & AC                                                                       & Classification                                                      & LR                                                                    & [10,195]                                                           & [0.05,5]            & $10^{-6}$                          \\ \cline{2-3} \cline{6-12} 
                                                                       & Wei et al. \cite{wei2021user}                        & 2021          &                                                              &                                                                       & Gaussian                                                                   & MA                                                                       & Classification                                                      & Shallow CNN                                                           & 50                                                                 & [4,20]              & $10^{-3}$                          \\ \cline{2-3} \cline{6-12} 
                                                                       & Liu et al. \cite{liu2021projected}                   & 2021          &                                                              &                                                                       & Gaussian                                                                   & GDP                                                                      & Classification                                                      & Shallow CNN                                                           & 100                                                                & [10,100]            & $10^{-3}$                          \\ \cline{2-3} \cline{6-12} 
                                                                       & Zheng et al. \cite{zheng2021federated}               & 2021          &                                                              &                                                                       & Gaussian                                                                   & GDP                                                                      & Classification                                                      & Shallow CNN                                                           & 100                                                                & [10,100]            & $10^{-3}$                          \\ \cline{2-3} \cline{6-12} 
                                                                       & Huang et al. \cite{huang2020dp}                      & 2020          &                                                              &                                                                       & Gaussian, Laplace                                                          & AC                                                                       & Classification                                                      & Shallow CNN                                                           & {10,100,1000}                                                      & [0.2,8]             & $[10^{-2},10^{-5}]$                \\ \cline{2-3} \cline{6-12} 
                                                                       & Wei et al. \cite{wei2020federated}                   & 2020          &                                                              &                                                                       & Gaussian                                                                   & MA                                                                       & Classification                                                      & Shallow CNN,LSTM                                                      & [10,20]                                                            & [0.12,2]            & $[10^{-2},10^{-5}]$                \\ \cline{2-3} \cline{6-12} 
                                                                       & Huang et al. \cite{huang2019dp}                      & 2019          &                                                              & \multirow{-16}{\multirowfit}{SL}                                      & Gaussian                                                                   & AC                                                                       & Regression                                                          & LR                                                                    & -                                                                  & [0.01,0.2]          & $[10^{-3},10^{-6}]$                \\ \cline{2-3} \cline{5-12} 
                                                                       & Yang et al. \cite{yang2023dynamic}                   & 2023          &                                                              &                                                                       & Gaussian                                                                   & RDP                                                                      & Classification                                                      & Shallow CNN                                                           & 50                                                                 & [2,16]              & $10^{-3}$                          \\ \cline{2-3} \cline{6-12} 
                                                                       & Xu et al. \cite{xu2023learning}                      & 2023          &                                                              &                                                                       & Gaussian                                                                   & RDP                                                                      & Classification                                                      & ResNet-50                                                             & [1262,9896000]                                                     & [10,20]             & $10^{-7}$                          \\ \cline{2-3} \cline{6-12} 
                                                                       & Shi et al. \cite{shi2023make}                        & 2023          &                                                              &                                                                       & Gaussian                                                                   & RDP                                                                      & Classification                                                      & ResNet-18                                                             & 500                                                                & [4,10]              & $\frac{1}{500}$                    \\ \cline{2-3} \cline{6-12} 
                                                                       & Zhang et al. \cite{zhang2022understanding}           & 2022          &                                                              &                                                                       & Gaussian                                                                   & MA                                                                       & Classification                                                      & Shallow CNN, ResNet-18                                                & 1920                                                               & [1.5,5]             & $10^{-5}$                          \\ \cline{2-3} \cline{6-12} 
                                                                       & Cheng et al. \cite{cheng2022differentially}          & 2022          &                                                              &                                                                       & Gaussian                                                                   & MA                                                                       & Classification                                                      & Shallow CNN, ResNet-18                                                & 3400                                                               & [2,8]               & $\frac{1}{3400}$                   \\ \cline{2-3} \cline{6-12} 
                                                                       & Bietti et al. \cite{bietti2022personalization}       & 2022          &                                                              &                                                                       & Gaussian                                                                   & MA                                                                       & Classification                                                      & Shallow CNN                                                           & 1000                                                               & [0.1,1000]          & $10^{-4}$                          \\ \cline{2-3} \cline{6-12} 
                                                                       & Andrew et al. \cite{andrew2021differentially}        & 2021          &                                                              &                                                                       & Gaussian                                                                   & RDP                                                                      & Classification                                                      & Shallow CNN                                                           & [500,342000]                                                       & [0.035,5]           & $[\frac{1}{500},\frac{1}{342000}]$ \\ \cline{2-3} \cline{6-12} 
                                                                       & Mcmahan et al. \cite{brendan2018learning}            & 2018          &                                                              &                                                                       & Gaussian                                                                   & MA                                                                       & Classification                                                      & LSTM                                                                  & [100,763430]                                                       & [2.0,4.6]           & $10^{-9}$                          \\ \cline{2-3} \cline{6-12} 
                                                                       & Geyer et al. \cite{geyer2017differentially}          & 2017          &                                                              & \multirow{-9}{\multirowfit}{CL}                                       & Gaussian                                                                   & MA                                                                       & Classification                                                      & Shallow CNN                                                           & {100, 1000, 10000}                                                 & 8                   & $[10^{-3},10^{-6}]$                \\ \cline{2-3} \cline{5-12} 
                                                                       & Chen et al. \cite{chen2022fundamental}               & 2022          &                                                              &                                                                       & Discrete Gaussian                                                          & RDP                                                                      & Classification                                                      & Shallow CNN                                                           & [100,1000]                                                         & [0,10]              & $10^{-2}$                          \\ \cline{2-3} \cline{6-12} 
                                                                       & Chen et al. \cite{chen2022poisson}                   & 2022          &                                                              &                                                                       & Poisson Binomial                                                           & RDP                                                                      & Classification                                                      & LR                                                                    & 1000                                                               & [0.5,6]             & $10^{-5}$                          \\ \cline{2-3} \cline{6-12} 
                                                                       & Wang et al. \cite{wang2020d2p}                       & 2020          &                                                              &                                                                       & Discrete Gaussian                                                          & RDP                                                                      & Classification                                                      & Shallow CNN                                                           & 100K                                                               & [2,4]               & $10^{-5}$                          \\ \cline{2-3} \cline{6-12} 
                                                                       & Stevens et al. \cite{stevens2022efficient}           & 2022          &                                                              &                                                                       & LWE                                                                        & RDP                                                                      & Classification                                                      & Shallow CNN                                                           & [500,1000]                                                         & [2,8]               & $10^{-5}$                          \\ \cline{2-3} \cline{6-12} 
                                                                       & Kairouz et al. \cite{kairouz2021distributed}         & 2021          &                                                              &                                                                       & Discrete Gaussian                                                          & zCDP                                                                     & Classification                                                      & Shallow CNN                                                           & 3400                                                               & [3,10]              & $\frac{1}{3400}$                   \\ \cline{2-3} \cline{6-12} 
                                                                       & Agarwal et al. \cite{agarwal2021skellam}             & 2021          &                                                              &                                                                       & Skellam                                                                    & RDP                                                                      & Classification                                                      & Shallow CNN                                                           & 1000k                                                              & [5,20]              & $10^{-6}$                          \\ \cline{2-3} \cline{6-12} 
                                                                       & Kerkouche et al. \cite{kerkouche2021compression}     & 2021          &                                                              &                                                                       & Gaussian                                                                   & MA                                                                       & Classification                                                      & Shallow CNN                                                           & [5011,6000]                                                        & [0.5,1]             & $10^{-5}$                          \\ \cline{2-3} \cline{6-12} 
                                                                       & Agarwal et al. \cite{agarwal2018cpsgd}               & 2018          &                                                              & \multirow{-8}{\multirowfit}{CL with SA}                               & Binomial                                                                   & AC                                                                       & Classification                                                      & LR                                                                    & 25M                                                                & [2,4]               & $10^{-9}$                          \\ \cline{2-3} \cline{5-12} 
                                                                       & Naseri et al. \cite{naseri2022local}                 & 2022          &                                                              & SL, CL                                                                & Gaussian                                                                   & RDP                                                                      & Classification                                                      & Shallow CNN,LSTM                                                      & [100,660120]                                                       & [1.2,10.7]          & $10^{-5}$                          \\ \cline{2-3} \cline{5-12} 
                                                                       & Yang et al. \cite{yang2023privatefl}                 & 2023          & \multirow{-35}{\multirowfit}{DP}                             & SL, CL, CL with SA                                                    & Gaussian, Skellam                                                          & RDP                                                                      & Classification                                                      & Shallow CNN                                                           & [40,500]                                                           & [2,8]               & $10^{-3}$                          \\ \cline{2-12} 
                                                                       & Triastcyn et al. \cite{triastcyn2019federated}       & 2019          & Bayesian DP                                                  & SL, CL                                                                & Gaussian                                                                   & RDP                                                                      & Classification                                                      & ResNet-50                                                             & [100,10000]                                                        & [0.2,4]             & $[10^{-3},10^{-6}]$                \\ \cline{2-12} 
                                                                       & Zhang et al. \cite{zhang2024dynamic}                 & 2024          &                                                              &                                                                       & Gaussian                                                                   & zCDP                                                                     & Classification                                                      & LR                                                                    & 20                                                                 & 1                   & $10^{-4}$                          \\ \cline{2-3} \cline{6-12} 
                                                                       & Varun et al. \cite{Varun24SSRFL}                     & 2024          &                                                              &                                                                       & SRR                                                                        & BC                                                                       & Classification                                                      & Shallow CNN                                                           & 100                                                                & [1,10]              & 0                                  \\ \cline{2-3} \cline{6-12} 
                                                                       & Zhang et al. \cite{zhang2023pfldp}                   & 2023          &                                                              &                                                                       & Gaussian                                                                   & AC                                                                       & Classification                                                      & LR, Shallow CNN                                                       & 100                                                                & [3,30]              & -                                  \\ \cline{2-3} \cline{6-12} 
                                                                       & Wang et al. \cite{wang2023ppefl}                     & 2023          &                                                              &                                                                       & EM, DMP-UE                                                                 & BC                                                                       & Classification                                                      & Shallow CNN                                                           & [10,50]                                                            & [0.1,1]             & 0                                  \\ \cline{2-3} \cline{6-12} 
                                                                       & Jiang et al. \cite{jiang2022signds}                  & 2023          &                                                              &                                                                       & EM                                                                         & BC                                                                       & Classification                                                      & Shallow CNN                                                           & [100,750]                                                          & [0.5,12]            & 0                                  \\ \cline{2-3} \cline{6-12} 
                                                                       & Li et al. \cite{li2023fedta}                         & 2023          &                                                              &                                                                       & Laplace                                                                    & BC                                                                       & Classification                                                      & Shallow CNN                                                           & 100                                                                & 78.5                & 0                                  \\ \cline{2-3} \cline{6-12} 
                                                                       & Lian et al. \cite{lian2022webfed}                    & 2022          &                                                              &                                                                       & Laplace                                                                    & BC                                                                       & Classification                                                      & Shallow CNN                                                           & 5                                                                  & [3,6]               & 0                                  \\ \cline{2-3} \cline{6-12} 
                                                                       & Mahawaga et al. \cite{mahawaga2022local}             & 2022          &                                                              &                                                                       & RAPPOR                                                                     & BC                                                                       & Classification                                                      & Shallow CNN                                                           & [2,100]                                                            & [0.5,10]            & 0                                  \\ \cline{2-3} \cline{6-12} 
                                                                       & Wang et al. \cite{wang2022safeguarding}              & 2022          &                                                              &                                                                       & RAPPOR                                                                     & BC                                                                       & Classification                                                      & LR                                                                    & [500,1800]                                                         & [0.1,10]            & 0                                  \\ \cline{2-3} \cline{6-12} 
                                                                       & Zhao et al. \cite{zhao2022privacy}                   & 2022          &                                                              &                                                                       & Adaptive-Harmony                                                           & BC                                                                       & Classification                                                      & Shallow CNN                                                           & 200                                                                & [1,10]              & 0                                  \\ \cline{2-3} \cline{6-12} 
                                                                       & Sun et al. \cite{sun2021ldp}                         & 2021          &                                                              &                                                                       & Adaptive-Duchi                                                             & BC                                                                       & Classification                                                      & Shallow CNN                                                           & [100,500]                                                          & [1,5]               & 0                                  \\ \cline{2-3} \cline{6-12} 
                                                                       & Yang et al. \cite{yang2021federated}                 & 2021          &                                                              &                                                                       & Laplace                                                                    & BC                                                                       & Classification                                                      & Shallow CNN                                                           & [200,1000]                                                         & [1,10]              & 0                                  \\ \cline{2-3} \cline{6-12} 
                                                                       & Wang et al. \cite{wang2020federated}                 & 2020          &                                                              &                                                                       & RRP                                                                        & AC                                                                       & Topic Modeling                                                      & LDA                                                                   & 150                                                                & [5,8]               & [0.05,0.5]                         \\ \cline{2-3} \cline{6-12} 
                                                                       & Zhao et al. \cite{zhao2020local}                     & 2020          &                                                              &                                                                       & Three output, PM-SUB                                                       & BC                                                                       & Classification                                                      & LR, SVM                                                               & 4M                                                                 & [0.5,4]             & 0                                  \\ \cline{2-3} \cline{6-12} 
                                                                       & Liu et al. \cite{liu2020fedsel}                      & 2020          &                                                              &                                                                       & RR, PM                                                                     & BC                                                                       & Classification                                                      & LR, SVM                                                               & 4W-10W                                                             & [0.5,16]            & 0                                  \\ \cline{2-3} \cline{6-12} 
                                                                       & Wang et al. \cite{wang2019collecting}                & 2019          & \multirow{-16}{\multirowfit}{LDP}                            & \multirow{-16}{\multirowfit}{-}                                       & PM                                                                         & BC                                                                       & Classification                                                      & LR, SVM                                                               & 4M                                                                 & [0.5,4]             & 0                                  \\ \cline{2-12} 
                                                                       & Truex et al. \cite{truex2020ldp}                     & 2020          & Condensed LDP                                                & -                                                                     & EM                                                                         & BC                                                                       & Classification                                                      & Shallow CNN                                                           & 50                                                                 & 1                   & 0                                  \\ \cline{2-12} 
                                                                       & Liu et al. \cite{liu2023echo}                        & 2023          &                                                              &                                                                       & Clipped-Laplace,Shuffle                                                    & AC                                                                       & Classification                                                      & LR                                                                    & 10000                                                              & 25.6                & $10^{-8}$                          \\ \cline{2-3} \cline{6-12} 
                                                                       & Liew et al. \cite{liew2023shuffled}                  & 2023          &                                                              &                                                                       & Harmony,Shuffle                                                            & RDP                                                                      & Classification                                                      & Shallow CNN                                                           & [50000,60000]                                                      & [2.8]               & -                                  \\ \cline{2-3} \cline{6-12} 
                                                                       & Liu et al. \cite{liu2021flame}                       & 2021          &                                                              & \multirow{-3}{\multirowfit}{CL}                                       & Laplace,Shuffle                                                            & BC, AC                                                                   & Classification                                                      & LR                                                                    & 1000                                                               & 4.696               & $5 \times 10^{-6}$                 \\ \cline{2-3} \cline{5-12} 
                                                                       & Chen et al. \cite{chen2024generalized}               & 2024          &                                                              &                                                                       & Duchi,Shuffle                                                              & GDP                                                                      & Classification                                                      & Shallow CNN                                                           & 100                                                                & [0.5, 100]          & $10^{-5}$                          \\ \cline{2-3} \cline{6-12} 
\multirow{-58}{\multirowfit}{Horizontal}                               & Girgis et al. \cite{girgis2021shuffled}              & 2021          & \multirow{-5}{\multirowfit}{Shuffle DP}                      & \multirow{-2}{\multirowfit}{SL}                                       & Laplace,Shuffle                                                            & AC                                                                       & Classification                                                      & Shallow CNN                                                           & 60000                                                              & [1,10]              & $10^{-5}$                          \\ \hline
                                                                       & Takahashi et al. \cite{takahashi2023eliminating}     & 2023          &                                                              &                                                                       & KRR                                                                        & BC                                                                       & Classification                                                      & GBDT                                                                  & 3                                                                  & [0.1,2.0]           & -                                  \\ \cline{2-3} \cline{6-12} 
                                                                       & Yang et al. \cite{yang2022differentially}            & 2022          & \multirow{-2}{\multirowfit}{Label DP}                        &                                                                       & Laplace, KRR                                                               & BC                                                                       & Classification                                                      & Shallow CNN                                                           & 2                                                                  & 1                   & 0                                  \\ \cline{2-4} \cline{6-12} 
                                                                       & Oh et al. \cite{oh2022differentially}                & 2022          &                                                              & \multirow{-3}{\multirowfit}{SL}                                       & Gaussian                                                                   & RDP                                                                      & Classification                                                      & VGG-16                                                                & 10                                                                 & [1,40]              & -                                  \\ \cline{2-3} \cline{5-12} 
                                                                       & Chen et al. \cite{chen2020vafl}                      & 2020          &                                                              &                                                                       & Gaussian                                                                   & GDP                                                                      & Classification                                                      & Shallow CNN                                                           & [3,8]                                                              & -                   & -                                  \\ \cline{2-3} \cline{6-12} 
                                                                       & Wang et al. \cite{wang2020hybrid}                    & 2020          &                                                              &                                                                       & Gaussian                                                                   & AC                                                                       & Classification                                                      & Shallow CNN                                                           & 2                                                                  & [0.001, 10]         & $10^{-2}$                          \\ \cline{2-3} \cline{6-12} 
                                                                       & Wu et al. \cite{wu2020privacy}                       & 2020          & \multirow{-4}{\multirowfit}{DP}                              & \multirow{-3}{\multirowfit}{CL}                                       & Laplace                                                                    & BC                                                                       & Classification                                                      & GBDT                                                                  & [2,10]                                                             & -                   & -                                  \\ \cline{2-12} 
                                                                       & Mao et al. \cite{mao2024secure}                      & 2024          &                                                              &                                                                       & Laplace, RR                                                                & BC                                                                       & Classification                                                      & Shallow CNN                                                           & 5                                                                  & [0.1,4.0]           & 0                                  \\ \cline{2-3} \cline{6-12} 
                                                                       & Tian et al. \cite{tian2024sf}                & 2024          & \multirow{-2}{\multirowfit}{LDP}                             & \multirow{-2}{\multirowfit}{-}                                        & RR                                                                         & BC                                                                       & Classification                                                      & GBDT                                                                  & 3                                                                  & 4                   & 0                                  \\ \cline{2-12} 
\multirow{-9}{\multirowfit}{Vertical}                                  & Li et al. \cite{li2022opboost}                       & 2022          & Condensed LDP                                                & -                                                                     & Discrete Laplace                                                           & BC                                                                       & Classification                                                      & GBDT                                                                  & 2                                                                  & [0.64,2.56]         & 0                                  \\ \hline
                                                                       & Wan et al. \cite{wan2023fedpdd}                      & 2023          &                                                              &                                                                       & Gaussian                                                                   & AC                                                                       & Recommendection                                                     & DeepFM                                                                & 2                                                                  & [0.05, 10]          & -                                  \\ \cline{2-3} \cline{6-12} 
                                                                       & Hoech et al. \cite{hoech2022fedauxfdp}               & 2022          &                                                              &                                                                       & Gaussian                                                                   & AC                                                                       & Classification                                                      & Resnet-18                                                             & 20                                                                 & [0.1,0.5]           & -                                  \\ \cline{2-3} \cline{6-12} 
                                                                       & Tian et al. \cite{tian2022seqpate}                   & 2022          &                                                              &                                                                       & Gaussian                                                                   & GDP                                                                      & Text Generation                                                     & GPT-2                                                                 & 2000                                                               & [3,5]               & $10^{-6}$                          \\ \cline{2-3} \cline{6-12} 
                                                                       & Sun et al. \cite{sun2021federated}                   & 2021          &                                                              &                                                                       & Random Sampling                                                            & AC                                                                       & Classification                                                      & Shallow CNN                                                           & 6                                                                  & [0.003,0.65]        & [0.006,0.65]                       \\ \cline{2-3} \cline{6-12} 
                                                                       & Papernot er al. \cite{papernot2018scalable}          & 2018          &                                                              &                                                                       & Gaussian                                                                   & RDP                                                                      & Classification                                                      & Resnet-18                                                             & 2                                                                  & [0.59,8.03]         & $10^{-8}$                          \\ \cline{2-3} \cline{6-12} 
                                                                       & Papernot er al. \cite{papernot2017semi}              & 2017          &                                                              & \multirow{-6}{\multirowfit}{SL}                                       & Laplace                                                                    & MA                                                                       & Classification                                                      & Shallow CNN                                                           & 2                                                                  & [2.04,8.19]         & $[10^{-5},10^{-6}]$                \\ \cline{2-3} \cline{5-12} 
                                                                       & Dodwadmath et al. \cite{dodwadmath2022preserving}    & 2022          &                                                              &                                                                       & Laplace                                                                    & MA                                                                       & Classification                                                      & Shallow CNN                                                           & 10                                                                 & [11.75,20]          & $10^{-5}$                          \\ \cline{2-3} \cline{6-12} 
                                                                       & Pan et al. \cite{pan2021fl}                          & 2021          & \multirow{-8}{\multirowfit}{DP}                              & \multirow{-2}{\multirowfit}{CL}                                       & Gaussian                                                                   & RDP                                                                      & Classification                                                      & Resnet-18                                                             & 100                                                                & [0.95,9.03]         & -                                  \\ \cline{2-12} 
\multirow{-9}{\multirowfit}{Transfer}                                 & Qi et al. \cite{qi2023differentially}                & 2023          & LDP                                                          & -                                                                     & KRR                                                                        & BC                                                                       & Classification                                                      & Shallow CNN                                                           & [2,5]                                                              & [2,7]               & 0                                  \\ \hline
\end{tabular}}
\begin{tablenotes}
	\footnotesize
	\item[1] 1. CM=Composition Mechanism, BC=Basic Sequential Composition Theory, AC=Advanced Sequential Composition Theory. 
  	\item[2] 2. LR=Logistic Regression, SVM=Support Vector Machine, GBDT=Gradient Boosting Decision Tree. 
\end{tablenotes}
\end{table*}

\subsection{Secure Multi-party Computation and Homomorphic Encryption in FL}\label{sec:section2.3}
Secure multi-party computation and homomorphic encryption are widely used cryptographic techniques for defending against privacy inference attacks in FL~\cite{li2020privacy,aono2017privacy,hao2019towards,bonawitz2017practical,xu2019verifynet,hao2019efficient,hardy2017private,chai2020secure,liu2020secure}. Li et al.~\cite{li2020privacy} proposed a privacy-preserving FL framework employing chained MPC to protect data privacy during collaborative learning among IoT devices. However, chained MPC requires complex cryptographic operations and introduces significant computational and communication overhead, limiting scalability in large IoT networks. Bonawitz et al.~\cite{bonawitz2017practical} introduced a practical secure aggregation protocol for FL, enabling a server to compute the sum of client-updated model parameters without accessing individual contributions. Nevertheless, the protocol requires careful synchronization across clients and is sensitive to user dropout, which can affect reliability and communication efficiency.

Aono et al.~\cite{aono2017privacy} proposed privacy-preserving deep learning using additively homomorphic encryption to allow secure computation of neural network functions on encrypted data. Although effective for privacy protection, this approach introduces considerable computational overhead and latency, making it unsuitable for real-time or large-scale applications. Hao et al.~\cite{hao2019towards} developed techniques to enhance federated deep learning efficiency and privacy by using model update sparsification, quantization, and secure aggregation. However, sparsification and quantization introduce additional complexity and may reduce model performance. Overall, cryptography-based approaches tend to incur high communication and computation costs due to the use of encryption or secret sharing.

\subsection{Recent Advances in DP+MPC for FL}\label{sec:section2.4}
Recent research on combining differential privacy with MPC in FL has explored integrated approaches to strengthen end-to-end privacy guarantees~\cite{xu2019hybridalpha,keller2024secure,zheng2024optimizing,chen2022fundamental,chen2022poisson,stevens2022efficient,kairouz2021distributed,agarwal2021skellam,kerkouche2021compression}. Xu et al.~\cite{xu2019hybridalpha} presented HybridAlpha, which combines federated learning with differential privacy and MPC to enhance privacy during collaborative model training across different entities. However, the combined use of DP and MPC increases both computation and communication overhead. Keller et al.~\cite{keller2024secure} proposed secure noise sampling within MPC to eliminate the need for clients to trust locally generated randomness, but this improvement comes at the cost of additional interaction steps and MPC computation. Zheng et al.~\cite{zheng2024optimizing} studied optimization techniques for the DP and MPC pipeline to improve the privacy-utility trade-off through coordinated mechanisms, although such coordination increases the complexity of system design and operation. Similarly, Chen et al.~\cite{chen2022fundamental} characterized the fundamental communication cost of secure aggregation for centrally differentially private federated learning and designed a near‑optimal scheme via sparse random projections that matches these bounds. However, achieving such guarantees still incurs substantial per‑client communication and additional computational overhead with careful parameter tuning, potentially limiting scalability in large‑scale deployments. 

To address the limitations described in Sections~\ref{sec:section2.2}, \ref{sec:section2.3}, and \ref{sec:section2.4}, we propose a novel privacy-preserving federated learning scheme with distributed differential privacy via secure aggregation, named DDP-SA. This scheme integrates local differential privacy with secure aggregation based on MPC, combining their respective strengths to defend against privacy inference attacks while maintaining acceptable model accuracy and efficiency. After detailed analysis, our DDP-SA framework emphasizes simplicity, scalability, and controllable linear cost through full-threshold additive secret sharing and client-side local DP.

\section{Preliminaries}\label{sec:section3}

\subsection{FedAvg Algorithm}
Federated learning (FL) is a distributed machine learning paradigm that enables multiple clients to collaboratively train a shared global model without centralizing their private datasets. It allows us to formally define the federated averaging (FedAvg) algorithm~\cite{mcmahan2017communication}, which serves as the foundation for our DDP-SA framework.

\textbf{Problem Setup.} Consider $n$ clients $\{C_1, C_2, \ldots, C_n\}$, where each client $C_i$ holds a private dataset $\mathcal{D}_i$ with $|\mathcal{D}_i| = N_i$ samples. The global objective is to minimize:
\begin{equation}
F(\theta) = \sum_{i=1}^{n} \frac{N_i}{N} F_i(\theta), \quad \text{where } F_i(\theta) = \frac{1}{N_i} \sum_{(x,y) \in \mathcal{D}_i} \ell(\theta; x, y),
\end{equation}
where $N = \sum_{i=1}^{n} N_i$ is the total number of samples, $\ell(\cdot)$ is the loss function, and $\theta$ denotes the model parameters.

\textbf{FedAvg Algorithm (Model Averaging Variant).} At communication round $t$:
\begin{enumerate}[leftmargin=*]
    \item \textbf{Server broadcast:} The parameter server sends the current global model $\theta^{(t)}$ to all clients.
    \item \textbf{Local updates:} Each client $C_i$ performs $E$ epochs of local SGD:
    \begin{equation}
    \theta_i^{(t+1)} = \theta^{(t)} - \eta \sum_{e=1}^{E} \nabla F_i(\theta_i^{(t,e)}),
    \end{equation}
    where $\eta$ is the learning rate and $\theta_i^{(t,e)}$ denotes client $i$'s model after local epoch $e$.
    \item \textbf{Server aggregation:} The parameter server computes the weighted average:
    \begin{equation}
    \theta^{(t+1)} = \sum_{i=1}^{n} \frac{N_i}{N} \theta_i^{(t+1)}.
    \end{equation}
\end{enumerate}

\textbf{Gradient Averaging Variant.} In this work, we focus on the gradient averaging variant in which clients send gradients rather than model parameters. At each round, client $C_i$ computes and sends:
\begin{equation}
g_i^{(t)} = \nabla F_i(\theta^{(t)}) = \frac{1}{N_i} \sum_{(x,y) \in \mathcal{D}_i} \nabla \ell(\theta^{(t)}; x, y).
\end{equation}
The server updates the global model as:
\begin{equation}
\theta^{(t+1)} = \theta^{(t)} - \eta \sum_{i=1}^{n} \frac{N_i}{N} g_i^{(t)}.
\end{equation}
This gradient-based formulation is equivalent to the original FedAvg algorithm~\cite{mcmahan2017communication} and serves as the target of our DDP-SA framework, where we apply local differential privacy and secure aggregation to protect $g_i^{(t)}$ during FL.

\subsection{Differential Privacy}
Differential privacy introduces randomness into a client's data or model updates before they are transmitted to the server to defend against privacy inference attacks in FL.

\subsubsection*{Definition 1 (Differential Privacy~\cite{dwork2014algorithmic})}
A randomized algorithm $\mathcal{M}$ with domain $\mathbb{N}^{|\mathcal{X}|}$ is $(\epsilon,\delta)$-differentially private if for all $\mathcal{S} \subseteq \text{Range}(\mathcal{M})$ and for all $x,y \in \mathbb{N}^{|\mathcal{X}|}$ such that $||x - y||_1 \leq 1$, 
\begin{equation}
\mathrm{Pr}[\mathcal{M}(x) \in \mathcal{S}] \leq e^{\epsilon} \, \mathrm{Pr}[\mathcal{M}(y) \in \mathcal{S}] + \delta,
\end{equation}
where $\epsilon$ defines the privacy budget and $\delta$ is the probability of privacy leakage. When $\delta = 0$, $\mathcal{M}$ is $\epsilon$-differentially private.

\subsubsection*{Definition 2 ($\ell_1$-Sensitivity~\cite{dwork2014algorithmic})}
The $\ell_1$-sensitivity of a function $f : \mathbb{N}^{|\mathcal{X}|} \rightarrow \mathbb{R}^k$ is:
\begin{equation}
\Delta f = \max_{\substack{x,y \in \mathbb{N}^{|\mathcal{X}|} \\ ||x - y||_1 = 1}} || f(x) - f(y) ||_1.
\end{equation}

\subsubsection*{Definition 3 (Laplace Distribution~\cite{dwork2014algorithmic})}
The Laplace distribution with scale $b$ has probability density function:
\begin{equation}
\mathrm{Lap}(x \mid b) = \frac{1}{2b} \exp\!\left(-\frac{|x|}{b}\right).
\end{equation}
Its variance is $\sigma^2 = 2b^2$.

\subsubsection*{Definition 4 (Laplace Mechanism~\cite{dwork2014algorithmic})}
Given any function $f : \mathbb{N}^{|\mathcal{X}|} \rightarrow \mathbb{R}^k$, the Laplace mechanism is:
\begin{equation}
\mathcal{M}_L(x, f(\cdot), \epsilon) = f(x) + (Y_1, \dots, Y_k),
\end{equation}
where $Y_i$ are i.i.d. random variables drawn from $\mathrm{Lap}(\Delta f / \epsilon)$.

\subsubsection*{Proposition 1 (Post-Processing~\cite{dwork2014algorithmic})}
If $\mathcal{M} : \mathbb{N}^{|\mathcal{X}|} \rightarrow R$ is $(\epsilon,\delta)$-differentially private and $f : R \rightarrow R'$ is any randomized mapping, then $f \circ \mathcal{M}: \mathbb{N} ^ {\left|\mathcal{X}\right|} \rightarrow \mathnormal{R}^{\prime}$ is also $(\epsilon,\delta)$-differentially private.

Differential privacy can be enforced without assuming trust in the central server by applying the mechanism $\mathcal{M}$ locally to each user's data before communication. This model, known as local differential privacy (LDP), is widely used in applications such as telemetry collection by Google, Apple, and Microsoft~\cite{erlingsson2014rappor,2017LearningWP,ding2017collecting}.

\subsection{Secure Multi-party Computation}\label{sec:section3.3}
Secure multi-party computation (MPC) enables mutually distrusting parties to collaboratively compute a function on their private inputs while ensuring that each party's input remains confidential~\cite{canetti1996adaptively}. In FL, MPC protects the privacy of client data by ensuring that only aggregated information is revealed. MPC can be instantiated through oblivious transfer, secret sharing, and threshold homomorphic encryption. In this paper, we focus on additive secret sharing (ASS), a full-threshold secret sharing scheme.

ASS splits a secret $S$ into $n$ shares $s_1, \dots, s_n$ in a finite field $\mathbb{Z}_p$ such that:
\begin{equation}
S \equiv \sum_{i=1}^{n} s_i \pmod{p}.
\end{equation}
Each share is uniformly random and reveals no information about $S$ on its own. The ASS procedure is presented in Algorithm~\ref{ASS Algorithm}.

\begin{algorithm}
\caption{ASS}
\label{ASS Algorithm}
\begin{algorithmic}[1]
\Statex \textbf{Input:} secret $S$, number of parties $n$
\Statex \textbf{Output:} shares $s_1, \dots, s_n$ such that $S \equiv \sum_{i=1}^{n} s_i \pmod{p}$
\State Choose a large prime $p$
\For{$i = 1$ to $n-1$}
    \State Sample $s_i$ uniformly at random from $\mathbb{Z}_p$
\EndFor
\State Compute $s_n \leftarrow S - \left(\sum_{i=1}^{n-1} s_i\right) \bmod p$
\For{each party $i = 1$ to $n$}
    \State Send share $s_i$ to party $i$
\EndFor
\Statex \textbf{Reconstruction:} When reconstruction is needed, all parties send their shares $s_1, \dots, s_n$ to the reconstructor, who computes
\Statex \hspace{\algorithmicindent} $S \leftarrow \left(\sum_{i=1}^{n} s_i\right) \bmod p$
\end{algorithmic}
\end{algorithm}

\textbf{Secure Aggregation in FL.} With ASS, each client shares its model updates across multiple intermediate servers. The servers aggregate shares locally and send aggregated shares to the parameter server, which reconstructs the global sum. No individual client update is ever revealed during this process.

\textbf{Security Interpretation.} Under the semi-honest model, any strict subset of additive shares is uniformly random and independent of the secret. Therefore, an adversary controlling fewer than all servers learns nothing about any individual client update, which aligns with classical MPC security definitions~\cite{canetti1996adaptively,bonawitz2017practical}.

\textbf{Quantization Error Bound.} Let $q(x)=\mathrm{round}(x \cdot \text{SF})/\text{SF}$ be fixed-point encoding with scaling factor $\text{SF}=10^{d_n}$. Then each coordinate satisfies $|q(x) - x| \le \tfrac{1}{2\, \text{SF}}$. If $p$ is chosen larger than the maximum possible aggregated magnitude, wrap-around in $\mathbb{Z}_p$ is avoided and the decoding error remains bounded by $\tfrac{1}{2\,\text{SF}}$, which is negligible for large $\text{SF}$ values.

\section{Methodology}\label{sec:section4}

The term ``Distributed DP'' refers to client-side local perturbation that achieves $(\epsilon,\delta)$-DP at the distributed client level, in contrast to central differential privacy. The term ``Secure Aggregation'' refers to full-threshold additive secret sharing (ASS), which protects noisy updates from being exposed during transmission or to the parameter server. Hence, DDP-SA stands for ``Distributed Differential Privacy via Secure Aggregation''. Unlike standalone LDP or CDP, and unlike schemes based solely on MPC, DDP-SA jointly provides statistical privacy through DP and cryptographic protection through ASS without degrading the original $(\epsilon,\delta)$ privacy guarantee. It also prevents the exposure of per-client updates during aggregation.

\begin{figure*}[t]
  \centering 
  \includegraphics[width=7.2in, trim=1.7cm 0.44cm 2.3cm 0.3cm, clip]{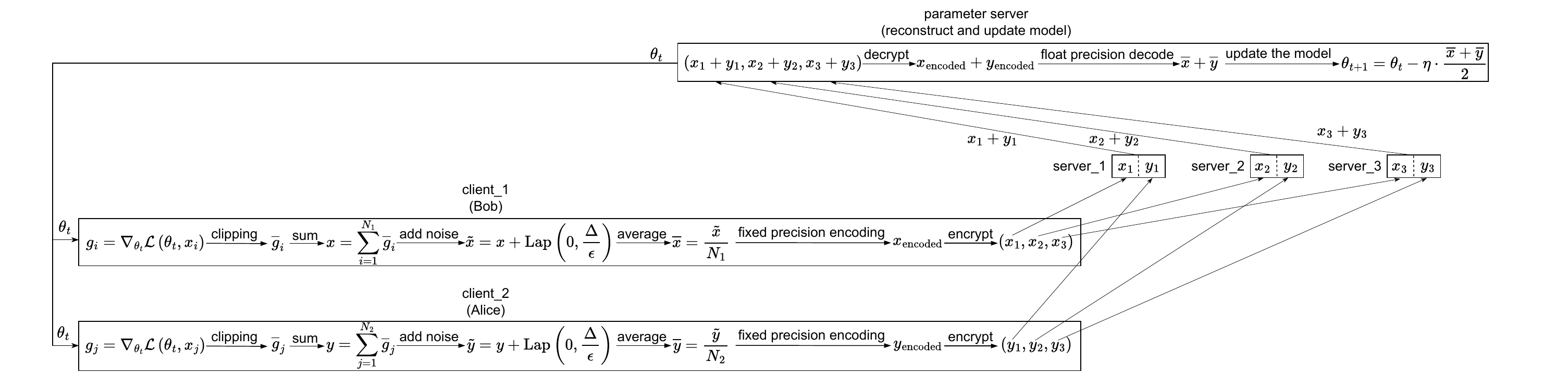}
  \caption{DDP-SA Framework diagram.}
  \label{Figure 1}
\end{figure*}

In this section, we provide a comprehensive overview of the DDP-SA architecture, which includes the system model and the threat model. Fig.~\ref{Figure 1} illustrates the schematic framework of DDP-SA. For simplicity, the figure considers two clients (Bob and Alice). Local gradients $x$ and $y$ are illustrated as scalars, and the encoded values $x_{\text{encoded}}$ and $y_{\text{encoded}}$ are divided into $m$ secret shares, to be sent to $m$ intermediate servers (in the illustration, $m=3$, although the protocol supports arbitrary $m$).

\subsection{System Model}
Our system model consists of three types of entities: clients, intermediate servers, and a parameter server. Compared to the conventional two-layer FL architecture with only clients and a parameter server, our design introduces a layer of intermediate servers that securely aggregate model updates using ASS. This layer ensures that the parameter server reconstructs only the aggregated update and not any single client's contribution. Each client trains a local neural network model through multiple rounds of iterative learning.

\textbf{Clients:} Each client holds a private dataset and has full control over its data. To prevent information leakage, a client computes its local gradient, adds Laplace noise, partitions the noisy gradient into multiple secret shares, and sends one share to each intermediate server. The client also receives global model parameters from the parameter server to compute its local gradient.

\textbf{Intermediate servers:} Each intermediate server possesses modest computational and storage capacity. It receives one secret share per client, performs addition operations on these shares to obtain a partial aggregate, and forwards this result to the parameter server.

\textbf{Roles of intermediate servers.} In addition to share aggregation, intermediate servers provide several system-level benefits:
\begin{enumerate}[leftmargin=*]
    \item \textbf{Share ingress and routing:} Receive per-server shares from all clients and route them reliably.
    \item \textbf{Batching and compression:} Combine shares in batches to improve network utilization.
    \item \textbf{Pipelined partial sums:} Forward intermediate sums upstream to reduce end-to-end latency.
    \item \textbf{Bandwidth offloading:} Replace $O(n \cdot d)$ client-to-server bandwidth with $O(m \cdot d)$ intermediate-to-server bandwidth.
    \item \textbf{Fault-domain isolation:} Reduce the effects of client churn and stragglers on the parameter server.
\end{enumerate}
Each intermediate server sees only a single additive share per client and performs simple addition in $\mathbb{Z}_p$, without ever decrypting a client's update.

\textbf{Parameter server:} The parameter server receives all aggregated partial sums from the intermediate servers, reconstructs the complete aggregated gradient, computes the average, and updates the global model accordingly. At the beginning of each training round, it broadcasts the updated model parameters to all clients. Because ASS is full-threshold, the parameter server can reconstruct the aggregate only after receiving all $m$ partial sums. Handling dropouts or applying dropout-tolerant aggregation techniques is outside the scope of this work and can be integrated as future improvements.

\textbf{Motivation for Additive Secret Sharing (ASS):}  
Although the architecture still includes a central parameter server, ASS enhances the end-to-end security of gradient transmission. Specifically, ASS prevents passive adversaries from learning perturbed gradients by observing communication links or compromising intermediate servers. The parameter server receives only aggregated results and, without collusion with all intermediate servers, cannot infer any single client's update. This limited use of MPC focuses solely on secure aggregation, rather than attempting full decentralization.

\subsection{Threat Model}\label{sec:section4.2}
We consider a semi-honest adversary model. All parties follow the protocol but may attempt to infer additional information from the data they observe. Our threat model includes the following assumptions.

\textbf{Adversary capability bounds:}
\begin{enumerate}[leftmargin=*]
    \item The adversary may corrupt at most $f$ intermediate servers, where $0 \le f < m$, and may collude with a bounded number of clients ($q_c$).
    \item The adversary may eavesdrop on a subset of communication links but cannot observe all links simultaneously.
    \item Communication channels are authenticated to prevent message tampering; confidentiality is provided by ASS rather than transport-layer encryption.
\end{enumerate}
Under these conditions, any strict subset of the $m$ shares is statistically independent of the secret, so adding the intermediate server layer does not increase privacy risk. Confidentiality fails only if an adversary controls all $m$ intermediate servers or if it observes all share-carrying links. This is a fundamental limitation of full-threshold ASS.

\textbf{Security goal:} Under the above assumptions, the confidentiality of each individual client's update is preserved. A strict subset of shares reveals no information about the client's perturbed gradient, and the parameter server learns only the aggregated update.

\textbf{Failure condition:} If an adversary controls all $m$ intermediate servers or simultaneously observes all share-carrying links, it can reconstruct the aggregated secret. This limitation is inherent to full-threshold ASS and consistent with secure aggregation literature~\cite{bonawitz2017practical}.

\textbf{Attacker placements considered:}
\begin{enumerate}[leftmargin=*]
    \item \textbf{External eavesdropper:} Can observe only a subset of communication links. Fewer than $m$ captured shares are insufficient to reconstruct any client's update.
    \item \textbf{Curious parameter server:} Sees only aggregated sums and cannot isolate any client's update unless colluding with all intermediate servers.
    \item \textbf{Corrupted intermediate servers (up to $f < m$):} Each observes only one share per client. A strict subset of shares leaks no information.
    \item \textbf{Curious clients:} Observe only the aggregated update, which prevents isolating the contribution of any other client.
\end{enumerate}

\textbf{Scope:} Active adversaries (for example message dropping, replay, or forging) are outside the scope of this work. Such attacks can be mitigated with standard authentication and robustness techniques. Handling large-scale dropouts is also outside our focus and can be incorporated with dropout-tolerant aggregation.

\subsection{DDP-SA}
\begin{figure*}[t]
    \centering
    \includegraphics[width=1.05\linewidth, trim=1cm 1cm 6cm 1cm, clip]{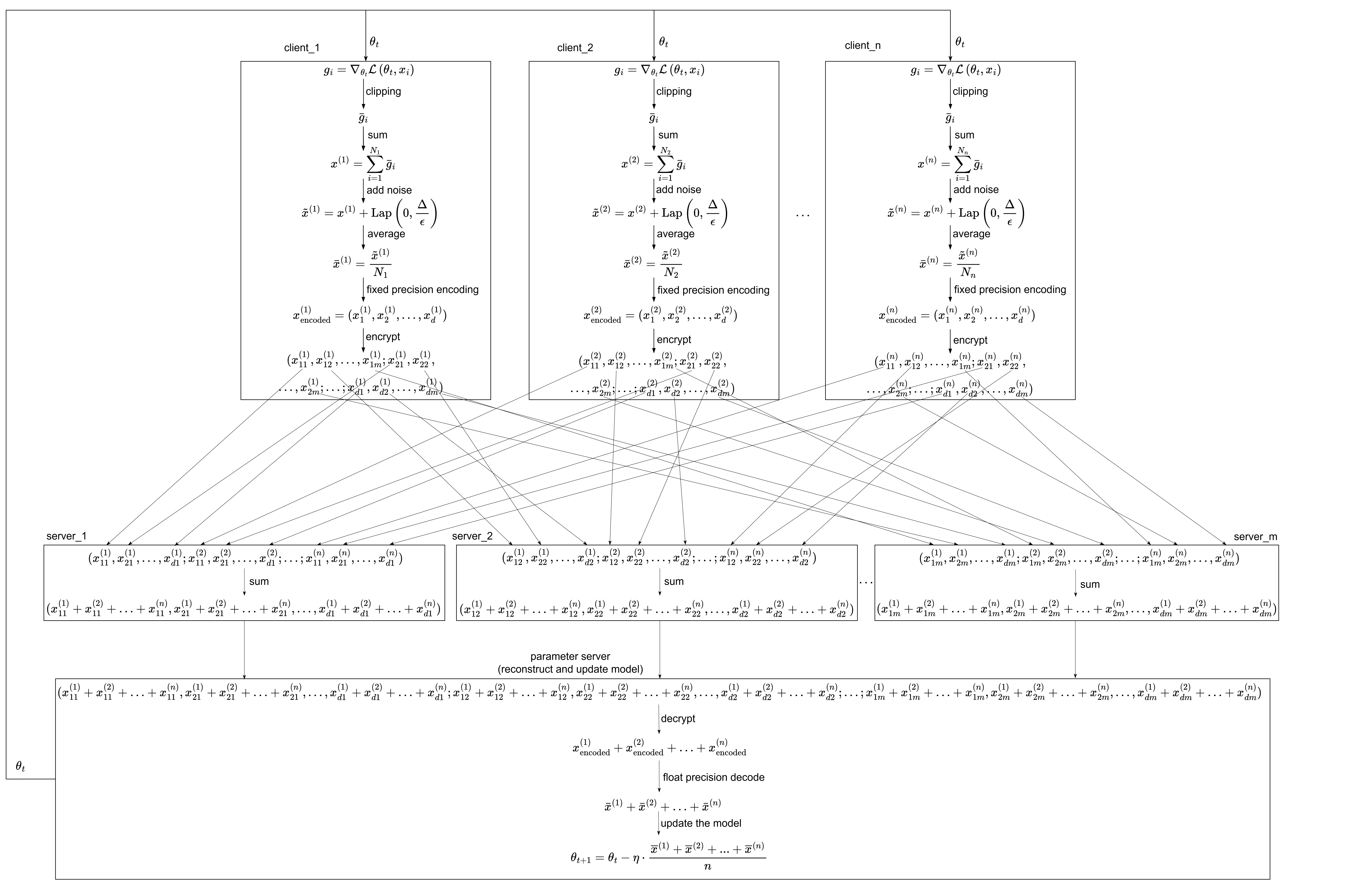}
    \caption{DDP-SA workflow, general scalable framework with $n$ clients, $m$ intermediate servers, and $d$-dimensional parameters.}
    \label{Figure 2}
\end{figure*}

The DDP-SA procedure is presented in Algorithm~\ref{DDP-SA Algorithm}. We consider $n$ clients and $m$ intermediate servers. Each client $C_i$ maintains a private dataset and a local model. Each intermediate server $S_j$ processes secret shares uploaded by clients. The algorithm proceeds as follows:
\begin{enumerate}[leftmargin=*]
    \item The parameter server broadcasts the initial model parameters to all clients.
    \item Each client computes the local gradient, adds Laplace noise, encodes the noisy gradient using fixed precision, generates secret shares, and uploads these shares to the intermediate servers.
    \item Each intermediate server aggregates secret shares from all clients and forwards the aggregated share to the parameter server.
    \item The parameter server reconstructs the complete aggregated gradient, updates the global model, and broadcasts updated parameters to all clients.
\end{enumerate}
This process repeats until the model converges or the maximum number of training rounds is reached. Fig.~\ref{Figure 2} shows the workflow of the DDP-SA framework. Each encoded gradient component is divided into $m$ shares, so each intermediate server receives exactly one share per component.

\textbf{Client-side operations:} Each client computes gradients for its samples, clips them using the $\ell_1$ norm, sums clipped gradients, adds Laplace noise, averages the noisy gradients, encodes the values using fixed precision, and partitions them into secret shares.

\textbf{Server-side operations:} Intermediate servers aggregate secret shares from all clients and send the aggregated results to the parameter server. The parameter server reconstructs the aggregated gradient and uses it to update the global model.

\textbf{Benefits of intermediate servers:}  
Intermediate servers improve scalability by reducing the parameter server's bandwidth load and enabling pipelined aggregation. Since each intermediate server receives only one share per client, no server can infer the client's update in isolation.

\section{Theoretical Privacy Analysis}\label{sec:section5}

\subsection{Single-Round Privacy Analysis}
In this section, we provide formal end-to-end privacy guarantees for the DDP-SA framework and analyze how privacy loss behaves when combining local differential privacy (LDP) with MPC-based secure aggregation.

\subsubsection*{Theorem 1 (End-to-end Privacy Guarantee)}\label{ddp-sa-privacy}
Let $\mathcal{M}_{DDP\text{-}SA}$ denote the DDP-SA mechanism in which each client applies an $(\epsilon, \delta)$-LDP mechanism to its local gradient before ASS-based secure aggregation. Then $\mathcal{M}_{DDP\text{-}SA}$ satisfies $(\epsilon, \delta)$-differential privacy end-to-end.

\paragraph*{Proof sketch}
The proof uses two observations. First, each client's local mechanism satisfies $(\epsilon, \delta)$-LDP by construction, since it is the Laplace mechanism with an appropriate noise scale. Second, the ASS-based secure aggregation is a deterministic post-processing of the noisy gradients. By the post-processing invariance of differential privacy~\cite{dwork2014algorithmic}, any deterministic function applied to differentially private outputs preserves the same privacy guarantee. Since the aggregation via ASS is deterministic given the noisy inputs, the end-to-end mechanism inherits the $(\epsilon, \delta)$-DP guarantee without degradation.
\hfill$\square$

\textbf{Privacy Loss Composition.} An important question is whether combining LDP with MPC introduces any additional privacy loss. The following result answers this.

\subsubsection*{Corollary 1 (No Additional Privacy Loss)}
The privacy budget of DDP-SA is equal to that of the underlying LDP mechanism. The secure aggregation via ASS introduces zero additional privacy loss.

\paragraph*{Proof sketch}
This holds because ASS provides information-theoretic security. Any strict subset of secret shares is uniformly random and independent of the underlying secret. Therefore, an adversary that observes only a subset of shares gains no additional information beyond what is already accounted for by the local DP guarantee.
\hfill$\square$

\textbf{Advantage over LDP Alone.} Although DDP-SA and standalone LDP provide the same formal $(\epsilon, \delta)$-DP guarantee, DDP-SA offers stronger protection in realistic adversarial settings:
\begin{itemize}
    \item \textbf{Communication security:} Individual client updates remain cryptographically protected during transmission, whereas LDP alone sends noisy gradients in plaintext.
    \item \textbf{Server-side protection:} The parameter server observes only aggregated updates, not individual client contributions, which provides an extra layer of protection beyond the DP noise.
    \item \textbf{Partial compromise resilience:} If an adversary compromises fewer than all $m$ intermediate servers, it learns nothing about individual client updates because of the information-theoretic security of ASS.
\end{itemize}

\textbf{Security Model.} The analysis assumes a semi-honest adversary model in which all parties follow the protocol but may attempt to infer private information from their views. Under this model, DDP-SA combines statistical privacy (from DP) with cryptographic privacy (from ASS), providing defense in depth against different attack vectors.

\subsection{Multi-Round Privacy Analysis}
For practical FL systems, it is essential to understand how privacy guarantees evolve over multiple training rounds. We now analyze the cumulative privacy loss when the DDP-SA mechanism is executed for $T$ training rounds.

\subsubsection*{Theorem 2 (Multi-Round Privacy Guarantee)}\label{thm:multi-round-privacy}
Let $\mathcal{M}_{DDP\text{-}SA}^{(T)}$ denote the DDP-SA mechanism running for $T$ rounds, where each round applies an $(\epsilon, \delta)$-LDP mechanism. Then:
\begin{enumerate}[leftmargin=*]
    \item \textbf{Basic composition:} $\mathcal{M}_{DDP\text{-}SA}^{(T)}$ satisfies $(T\epsilon, T\delta)$-differential privacy.
    \item \textbf{Advanced composition:} For any $\delta' > 0$, $\mathcal{M}_{DDP\text{-}SA}^{(T)}$ satisfies $(\epsilon_{\text{total}}, \delta_{\text{total}})$-differential privacy, where
    \begin{equation}
        \epsilon_{\text{total}} = \epsilon\sqrt{2 T \ln(1/\delta')} + \epsilon T (e^{\epsilon} - 1), \quad 
        \delta_{\text{total}} = T\delta + \delta'.
    \end{equation}
\end{enumerate}

\paragraph*{Proof sketch}
By \hyperref[ddp-sa-privacy]{Theorem 1}, each individual round of DDP-SA satisfies $(\epsilon, \delta)$-DP. Applying the standard composition theorems for differential privacy~\cite{dwork2014algorithmic} to the sequence of $T$ rounds yields the stated bounds. The basic composition theorem yields $(T\epsilon, T\delta)$-DP. The advanced composition theorem gives a significantly tighter bound for $\epsilon_{\text{total}}$ when $T$ is large. For example, if $\epsilon = 0.1$, $T = 1000$, and $\delta' = 10^{-4}$, then the advanced composition bound gives $\epsilon_{\text{total}} \approx 24.09$, whereas the basic composition bound gives $\epsilon_{\text{total}} = 100$. The latter corresponds to a much larger privacy loss. Hence, advanced composition is preferable for long-running federated learning systems.
\hfill$\square$

\textbf{Privacy Budget Allocation Strategies.} To manage cumulative privacy loss over multiple rounds, we consider two allocation strategies for the privacy budget:
\begin{enumerate}[leftmargin=*]
    \item \textbf{Uniform allocation:} Divide a total budget $\epsilon_{\text{total}}$ equally across $T$ rounds, that is, $\epsilon_{\text{per-round}} = \epsilon_{\text{total}}/T$.
    \item \textbf{Adaptive allocation:} Allocate more budget to early rounds, when gradients tend to have larger magnitude, using exponential decay,
    \begin{equation}
        \epsilon_t = \epsilon_{\text{total}} \cdot \frac{\alpha^{t-1}}{\sum_{i=1}^T \alpha^{i-1}}, \quad \alpha \in (0,1).
    \end{equation}
\end{enumerate}

\textbf{Comparison with Multi-Round LDP.} Both DDP-SA and a pure LDP approach experience the same formal composition of DP parameters over multiple rounds, since they apply the same per-round DP mechanism. However, DDP-SA maintains additional protections, such as encrypted communication and server-side protection, in every round. As a result, DDP-SA offers stronger practical protection than LDP alone, even though the formal $(\epsilon, \delta)$ parameters are identical.

\textbf{Practical Implications.} As shown in \hyperref[thm:multi-round-privacy]{Theorem 2}, the cumulative privacy loss of DDP-SA can be bounded under both basic and advanced composition. For long-running FL systems, practitioners must carefully choose the total privacy budget and its allocation across rounds. The DDP-SA framework supports such budget management while retaining the cryptographic protections of secure aggregation throughout the entire training process.

\begin{algorithm}
\caption{DDP-SA}
\label{DDP-SA Algorithm}  
\begin{algorithmic}[1]
\Statex \textbf{Input:} set of clients $C = \{C_1, C_2, \ldots, C_n\}$, number of training rounds $T$, global model parameters $\theta$, set of intermediate servers $S = \{S_1, S_2, \ldots, S_m\}$, privacy budget $\epsilon$, clipping threshold $\Delta$ for the $\ell_1$ norm, number of samples $N_i$ for client $C_i$, learning rate $\eta$, total number of samples $N$ across all clients, large prime $p$, loss function $\mathcal{L}$, gradient $\nabla_{\theta_t} \mathcal{L}(\theta_t, x_j)$, fixed precision scaling factor $\text{SF}$, number of decimal places $d_n$ to preserve
\Statex \textbf{Output:} trained global model $\theta_T$
\For {each round $t = 0, 1, \dots, T - 1$}
    \State Parameter server broadcasts current model parameters $\theta_t$ to all clients
    \For{each client $C_i$ in parallel}
        \State $\nabla \theta \leftarrow 0$
        \For{each sample $x_j$ in $C_i$'s local dataset}
            \State $\mathbf{g}_t(x_j) \leftarrow \nabla_{\theta_t} \mathcal{L}(\theta_t, x_j)$
            \State $\overline{\mathbf{g}}_t(x_j) \leftarrow \mathbf{g}_t(x_j) \big/ \max\!\left(1, \frac{\|\mathbf{g}_t(x_j)\|_1}{\Delta}\right)$
            \State $\nabla \theta \leftarrow \nabla \theta + \overline{\mathbf{g}}_t(x_j)$
        \EndFor
        \State $\tilde{\mathbf{g}}_t \leftarrow \frac{1}{N_i}\left(\nabla \theta + \mathrm{Lap}\!\left(0, \frac{\Delta}{\epsilon}\right)\right)$
        \State $\text{SF} \leftarrow 10^{d_n}$
        \State $\tilde{\mathbf{g}}_{t,\text{encoded}} \leftarrow \mathrm{round}(\tilde{\mathbf{g}}_t \times \text{SF})$
        \State $\text{shares} \leftarrow C_i.\text{secret\_share}(\tilde{\mathbf{g}}_{t,\text{encoded}}, S)$
        \For{each share $\text{shares}[j]$}
            \State send $\text{shares}[j]$ to $S_j$
        \EndFor
    \EndFor
    \For{each server $S_j$ in parallel}
        \State $\nabla \theta_{\text{agg},j} \leftarrow \text{sum of shares from all clients at } S_j$
        \State send $\nabla \theta_{\text{agg},j}$ to the parameter server
    \EndFor
    \State $\nabla \theta_{\text{agg}} \leftarrow \left(\sum_{j=1}^m \nabla \theta_{\text{agg},j}\right) \bmod p$
    \State $\nabla \theta_{\text{agg}} \leftarrow \nabla \theta_{\text{agg}} \big/ \text{SF}$
    \State $\theta_{t+1} \leftarrow \theta_t - \eta \cdot \frac{N_i}{N} \cdot \nabla \theta_{\text{agg}}$
\EndFor
\State \Return $\theta_{T}$
\end{algorithmic}
\end{algorithm}

\section{Experimental Evaluation}\label{sec:section6}
In this section, we present extensive experiments that verify the proposed DDP-SA scheme. The evaluation covers efficiency, accuracy, privacy, and detailed performance analysis.

\subsection{Experimental Setup}
Python, PyTorch 1.4.0, and PySyft 0.2.9 were used to implement and evaluate the proposed scheme. All experiments were conducted on GitHub Codespaces equipped with 16 CPU cores, 64 GB RAM, and 128 GB of storage.

A synthetic dataset was created by generating a $10000 \times 2$ array of random samples from a uniform distribution. For each row, the two values were summed and the constant 1 was added to obtain the corresponding label. The learning task is therefore a simple linear regression of the form $y = x_1 + x_2 + 1$. The data were split into training, validation, and test sets using a ratio of 60 percent, 20 percent, and 20 percent, respectively, and the training data were distributed evenly among all clients. Because all samples come from the same distribution, only the independent and identically distributed (IID) case is considered.

A two-layer neural network was used for fitting, with two neurons in the input layer and one neuron in the output layer. For the No-Private mechanism (which uses neither MPC nor LDP) and the MPC mechanism, standard SGD with learning rate 0.1 was used. For the LDP and DDP-SA mechanisms, the Adam optimizer with learning rate 0.001 was used. In both LDP and DDP-SA, each client clips per-sample gradients with the same $\ell_1$ threshold $\Delta$, sums the clipped gradients, adds IID Laplace noise with scale $\Delta/\epsilon$, and averages the noisy gradients locally before transmission and encoding. All optimizers and hyperparameters were identical across LDP and DDP-SA. The privacy budget $\epsilon$ was set to 0.1. The sensitivity $\Delta$ was chosen as the median of the $\ell_1$ norms of the unclipped gradients across training. The number of retained decimal places $d_n$ was set to 10.

Because the differential privacy mechanism is stochastic, each reported result is averaged over multiple runs. In addition, reconstruction at the parameter server requires receipt of all aggregated results from the intermediate servers.

\subsection{Efficiency Analysis}
The efficiency analysis of the DDP-SA scheme focuses on two metrics: communication cost and computational cost. Communication cost is evaluated from the parameter server’s perspective and includes communication between the parameter server and clients, as well as between intermediate servers and the parameter server. Communication between clients and intermediate servers is excluded unless stated otherwise.

\begin{figure}[t]
  \centering
  \includegraphics[width=3.2in]{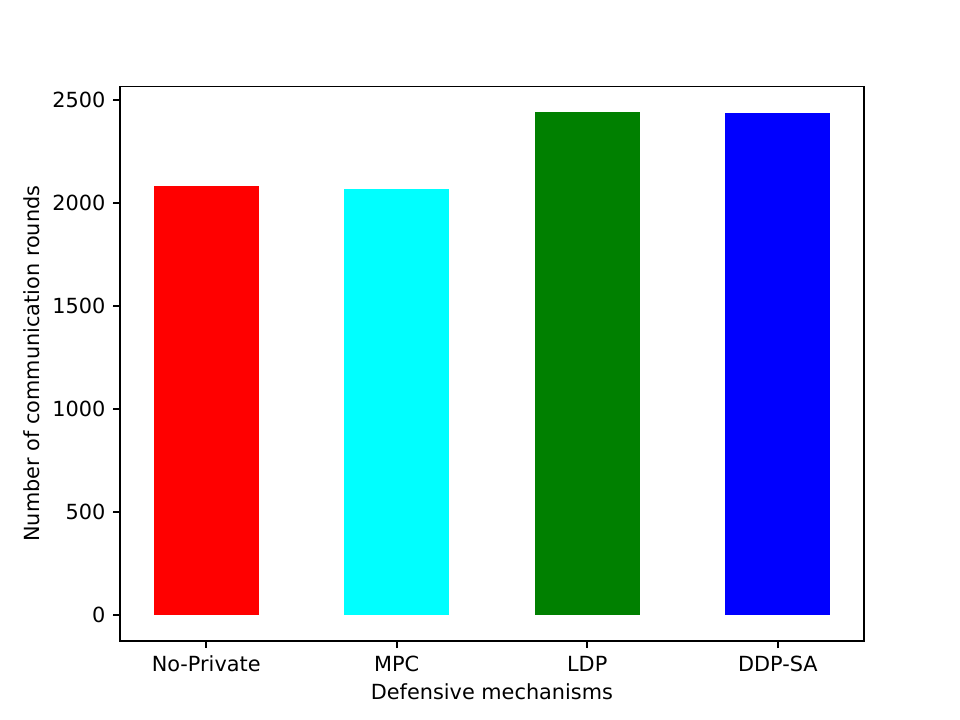}
  \caption{Number of communication rounds for different defensive mechanisms.}
  \label{Figure 3}
\end{figure}

Fig.~\ref{Figure 3} reports the total number of communication rounds until convergence under different defensive mechanisms. The No-Private and LDP mechanisms require 2082 and 2444 rounds, respectively. The MPC and DDP-SA mechanisms require 2070 and 2436 rounds, respectively. The results show that MPC behaves similarly to No-Private because neither mechanism introduces local noise and both use SGD with learning rate 0.1. Likewise, DDP-SA behaves similarly to LDP because both use local noise, clipping, and the Adam optimizer with learning rate 0.001. Optimizer choice can also influence round counts.

\begin{figure}[t]
  \centering
  \includegraphics[width=3.2in]{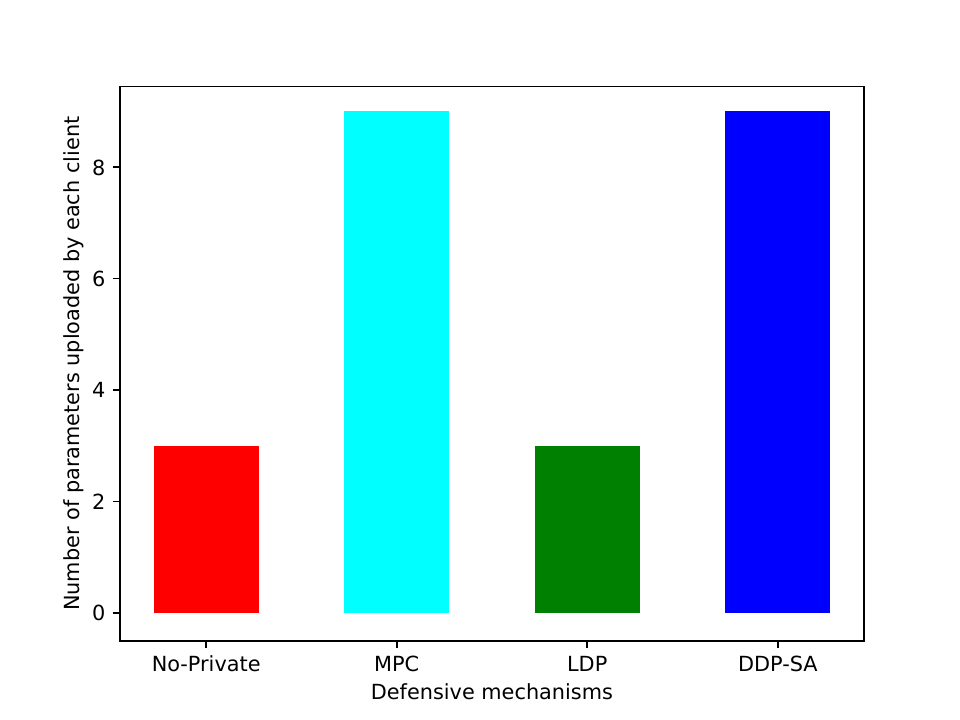}
  \caption{Number of parameters uploaded per client for different defensive mechanisms. Results shown for $m=3$.}
  \label{Figure 4}
\end{figure}

Fig.~\ref{Figure 4} shows the number of parameters uploaded per client under each mechanism. Both No-Private and LDP upload 3 parameters (model dimension $d = 3$). MPC and DDP-SA upload $3m$ parameters because each gradient component is split into $m$ secret shares. In our experiments, $m = 3$ was chosen as a practical trade-off between security and cost. The protocol supports arbitrary values of $m$, communication cost scales linearly with $m$, and confidentiality holds unless all $m$ paths are compromised.

\begin{figure}[t]
  \centering
  \includegraphics[width=3.2in]{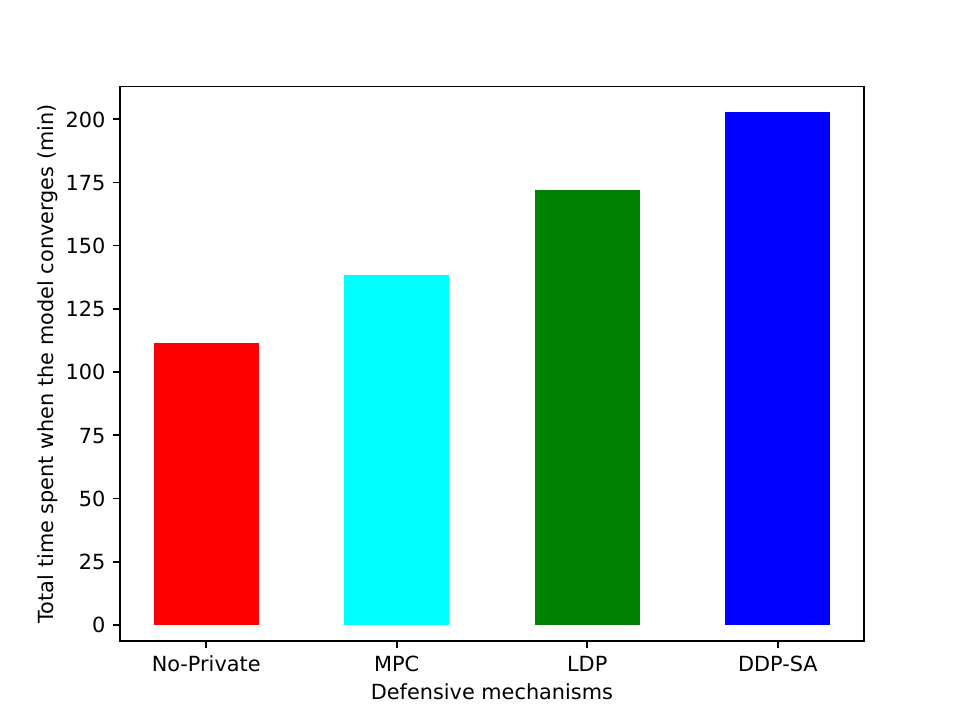}
  \caption{Total time to convergence for different defensive mechanisms.}
  \label{Figure 5}
\end{figure}

\begin{figure}[t]
  \centering
  \includegraphics[width=3.2in]{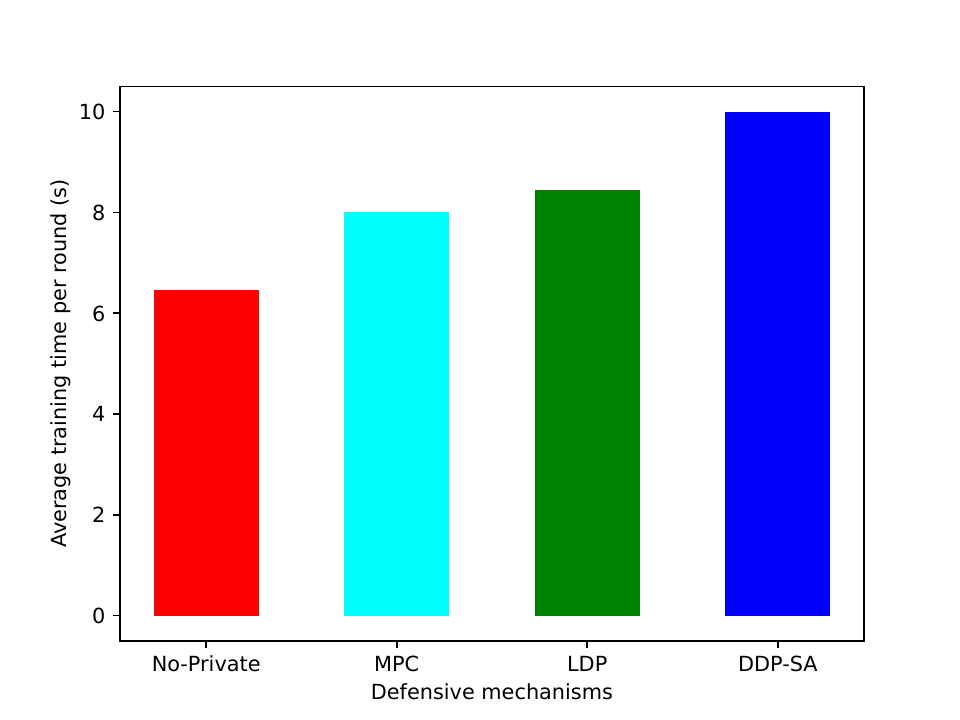}
  \caption{Average training time per round for different defensive mechanisms.}
  \label{Figure 6}
\end{figure}

Fig.~\ref{Figure 5} shows the total time to convergence for each mechanism. The No-Private and MPC mechanisms take 112 minutes and 138 minutes, respectively. The LDP and DDP-SA mechanisms take 172 minutes and 203 minutes, respectively. Fig.~\ref{Figure 6} shows the average training time per round. No-Private and MPC require 6.4553 seconds and 8 seconds per round, while LDP and DDP-SA require 8.4452 seconds and 10 seconds per round. 

From these results, we conclude that DDP-SA incurs slightly higher communication and computation overhead than LDP or MPC. However, the overhead remains acceptable and controllable for practical settings.

\subsection{Detailed Component-wise Overhead Analysis}
We now present a quantitative breakdown of computational and communication overhead to isolate the contributions of LDP and MPC.

\textbf{Computational Overhead Breakdown.} Table~\ref{tab:computation-breakdown} summarizes the per-client per-round computation cost:
\begin{itemize}
    \item \textbf{LDP overhead:} Accounts for 92.12 percent of total computation, dominated by gradient clipping. This cost scales linearly with the parameter dimension $d$.
    \item \textbf{MPC overhead:} Accounts for 7.88 percent of total computation, dominated by share transmission which scales as $O(d \cdot m)$.
    \item \textbf{Combined DDP-SA overhead:} Dominated by gradient clipping with scaling $O(d)$.
    \item \textbf{Primary bottleneck:} Gradient clipping rather than cryptographic operations.
    \item \textbf{Server-side operations excluded:} Aggregation and reconstruction at intermediate servers and the parameter server are not part of the client overhead.
\end{itemize}

\begin{table*}[t]
\caption{Computational Overhead Breakdown per Client per Round}
\label{tab:computation-breakdown}
\centering
\begin{tabular}{|c|c|c|c|c|}
\hline
Component & Operation & Time (ms) & Percentage of Total & Scalability \\
\hline
\multirow{3}{\multirowfit}{LDP} 
& Gradient Clipping & 547.95 & 92.07\% & $O(d)$ \\
& Noise Generation & 0.24 & 0.04\% & $O(d)$ \\
& Noise Addition & 0.05 & 0.01\% & $O(d)$ \\
\hline
\multirow{3}{\multirowfit}{MPC} 
& Fixed-Point Encoding & 0.22 & 0.04\% & $O(d)$ \\
& Secret Sharing & 0.53 & 0.09\% & $O(d \cdot m)$ \\
& Share Transmission & 46.13 & 7.75\% & $O(d \cdot m)$ \\
\hline
DDP-SA & All Operations & 595.12 & 100.0\% & $O(d \cdot m)$ \\
\hline
\end{tabular}
\end{table*}

\textbf{Communication Overhead Breakdown.} Table~\ref{tab:communication-breakdown} reports detailed bandwidth usage:
\begin{itemize}
    \item \textbf{LDP:} No additional overhead relative to No-Private.
    \item \textbf{MPC:} Uploads $m$ shares per gradient component, giving a factor of $m$ overhead.
    \item \textbf{DDP-SA:} Identical to MPC for communication overhead.
    \item \textbf{Intermediate server communication:} Adds $4d \cdot m$ bytes to the system but has no effect on clients.
\end{itemize}

\begin{table*}[t]
\caption{Communication Overhead Breakdown per Client per Round}
\label{tab:communication-breakdown}
\centering
\begin{tabular}{|c|c|c|c|c|}
\hline
Component & Direction & Bytes & Percentage of Total & Scalability \\
\hline
\multirow{2}{\multirowfit}{LDP} 
& PS to Client & $4d$ & 50.0\% & $O(d)$ \\
& Client to PS & $4d$ & 50.0\% & $O(d)$ \\
\hline
\multirow{3}{\multirowfit}{MPC} 
& PS to Client & $4d$ & 25.0\% & $O(d)$ \\
& Client to Intermediate Servers (excluded) & $4d \cdot m$ & - & $O(d \cdot m)$ \\
& Intermediate Servers to PS & $4d \cdot m$ & 75.0\% (for $m=3$) & $O(d \cdot m)$ \\
\hline
\multirow{3}{\multirowfit}{DDP-SA}
& PS to Client & $4d$ & 25.0\% & $O(d)$ \\
& Client to Intermediate Servers (excluded) & $4d \cdot m$ & - & $O(d \cdot m)$ \\
& Intermediate Servers to PS & $4d \cdot m$ & 75.0\% (for $m=3$) & $O(d \cdot m)$ \\
\hline
\end{tabular}
\end{table*}

\textbf{Scalability Analysis.} Both tables show how overhead scales with system parameters:
\begin{itemize}
    \item \textbf{Parameter dimension $d$:} All methods scale linearly with $d$.
    \item \textbf{Number of intermediate servers $m$:} LDP unaffected. MPC and DDP-SA scale linearly with $m$.
    \item \textbf{Number of clients $n$:} Per-client cost unchanged. Total system overhead grows linearly in $n$.
\end{itemize}

From the scalability analysis, we can see that the DDP-SA improves scalability compared to LDP: it converts $n$ client uplinks into $m$ intermediate server uplinks (with $m\!\ll\!n$), reduces the parameter server's per-round ingress bandwidth from $4nd$ to $4md$, which enables scalability to many clients and long training horizons.

\begin{figure*}[!t]
\centering
\subfloat[]{\includegraphics[width=2.95in]{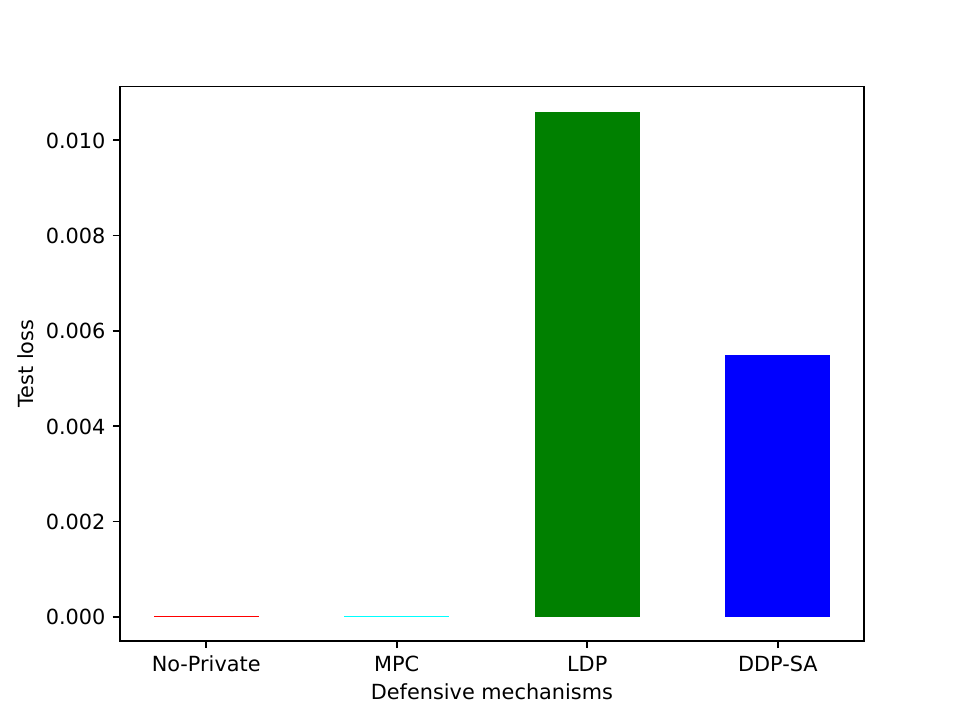}%
}
\hfil
\subfloat[]{\includegraphics[width=2.95in]{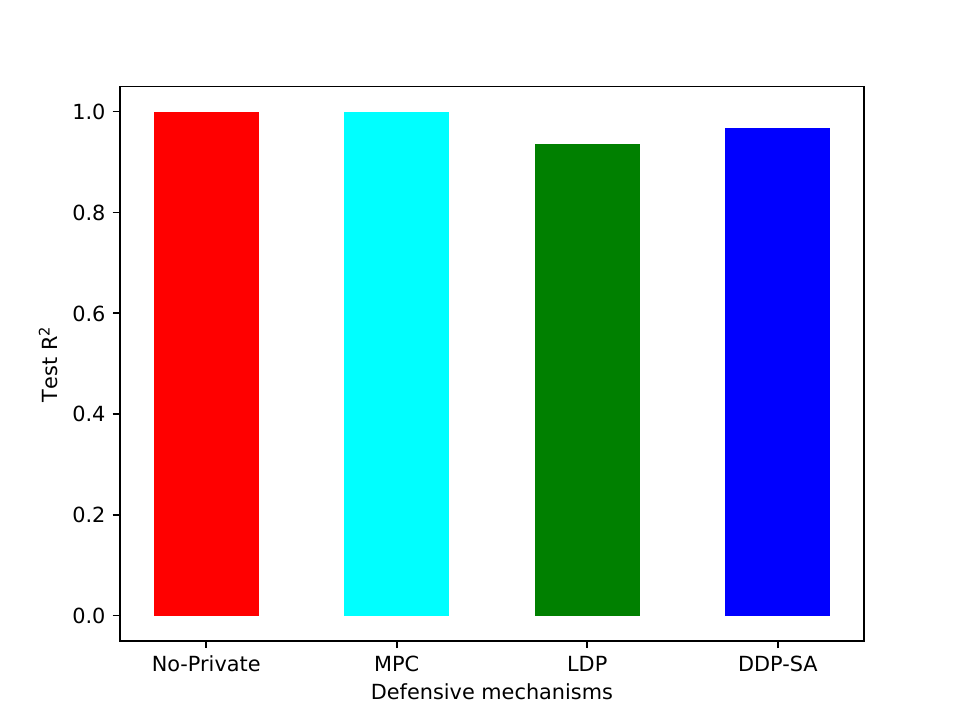}%
}
\caption{Accuracy for different defensive mechanisms. (a) Test loss. (b) Test $\text{R}^2$.}
\label{Figure 7}
\end{figure*}

\subsection{Accuracy Analysis}
We use test loss and test $\text{R}^2$ (coefficient of determination) to evaluate the accuracy of the trained global model. Fig.~\ref{Figure 7}(a) shows the test loss under different defensive mechanisms. As shown in Fig.~\ref{Figure 7}(a), the test loss of No-Private and MPC is close to zero (around $10^{-12}$), while the test loss of LDP and DDP-SA is 0.0106 and 0.0055, respectively. Thus, the test loss of LDP and DDP-SA is slightly higher than that of No-Private and MPC.

Fig.~\ref{Figure 7}(b) shows the test $\text{R}^2$ for different mechanisms. The test $\text{R}^2$ of both No-Private and MPC is 0.9999, while the test $\text{R}^2$ of LDP and DDP-SA is 0.9357 and 0.9666, respectively. Hence, LDP and DDP-SA exhibit slightly lower test $\text{R}^2$ than No-Private and MPC, while DDP-SA achieves a higher test $\text{R}^2$ than LDP.

From these results, we conclude that DDP-SA incurs some accuracy loss relative to No-Private and MPC, but the loss is acceptable and controllable. Moreover, Fig.~\ref{Figure 7} shows that MPC and No-Private achieve essentially identical test loss and test $\text{R}^2$, which indicates that the MPC computation and fixed-point encoding are effectively lossless. This confirms that the choice $d_n = 10$ is appropriate and is consistent with the negligible decoding error for large scaling factors $\text{SF}$ discussed in Section~\ref{sec:section3.3}. 

\subsection{Empirical Privacy Evaluation}

\subsubsection{Analysis of Privacy Protection Strength}
The privacy budget $\epsilon$ quantifies the privacy protection strength. Smaller values of $\epsilon$ provide stronger privacy. Fig.~\ref{Figure 8}(b) shows the effect of different values of $\epsilon$ on the test $\text{R}^2$. As $\epsilon$ increases, the test $\text{R}^2$ of both DDP-SA and LDP increases, but DDP-SA consistently achieves higher $\text{R}^2$ than LDP. Hence, for a fixed target accuracy, DDP-SA can operate with a smaller privacy budget than LDP, which means that DDP-SA achieves stronger privacy protection.

MPC can be viewed as a special case of DDP-SA where the privacy budget is effectively infinite (no noise is added to local gradients) and the clipping norm is set to the maximum gradient norm (clipping has no practical effect). In this sense, DDP-SA can also provide stronger privacy protection than pure MPC. The same conclusion can be drawn from Fig.~\ref{Figure 8}(a).

\subsubsection{Analysis of Privacy Leakage}
\subsubsection*{Lemma 1 (Strict-subset Indistinguishability)}
    Let $S$ be a client's (noisy) update and let $\{s_1,\dots,s_m\}$ be its ASS shares over $\mathbb{Z}_p$. 
    For any strict subset $K\subset\{1,\dots,m\}$, 
    \begin{equation}
    I\!\left(S;\{s_k\}_{k\in K}\right)=0.
    \end{equation}
    Consequently, if each intermediate server (or link) is independently compromised with probability $q$, then the probability of reconstructing $S$ is $q^m$, which decreases exponentially in $m$.

We now analyze privacy leakage for MPC, LDP, and DDP-SA using the DDP-SA workflow.

\begin{enumerate}[leftmargin=*]
    \item \textbf{MPC:} As discussed in Section~\ref{sec:section4.2}, an external adversary can attempt to intercept communication among clients, intermediate servers, and the parameter server. When eavesdropping on client to intermediate server communication, the adversary sees only a single secret share in $\mathbb{Z}_p$, which is uniformly random and independent of the secret, so any strict subset of shares is information-theoretically useless. Interception between intermediate servers and the parameter server reveals only the sum of secret shares, which does not expose any single client update. From the parameter server to the clients, the adversary can observe only global model parameters, which aggregate updates from many clients and do not reveal individual inputs.

    The parameter server receives sums of secret shares and reconstructs the complete gradient, but secure aggregation prevents it from isolating any individual client's gradient. An intermediate server receives only one share per client and cannot reconstruct the gradient. A local client can access global model parameters. In a two-client scenario, a client could infer the other client's gradient from the difference between the global and its own gradient, which can reveal private information. However, with more than two clients, only aggregated gradients are available, which obscure individual contributions.

    \item \textbf{LDP:} If an adversary eavesdrops on communication between a client and the parameter server, it observes only locally perturbed gradients. The adversary cannot recover the exact original data due to the noise, although, depending on the noise level, some limited inference may be possible. The parameter server receives only perturbed gradients and aggregate statistics based on them. It cannot deduce precise information about any individual update. Because LDP is applied locally before any sharing, no client or server can reverse the perturbation and recover the original data. Any further computation or model training on these noisy gradients preserves the DP guarantees by the post-processing property.

    \item \textbf{DDP-SA:} For DDP-SA, if an adversary eavesdrops on client to intermediate server communication, it observes only a single secret share per client, which is uniformly random and independent of the underlying noisy update. Thus, no information can be inferred from any strict subset of shares. If the adversary intercepts traffic between intermediate servers and the parameter server, it observes only partial sums of shares which do not reveal individual contributions. Observing communication from the parameter server to the clients allows access only to the global model parameters, which are functions of the locally perturbed gradients. Due to LDP and the post-processing invariance of DP, these global parameters do not leak additional information beyond what is already permitted by the DP guarantee.

    The parameter server can reconstruct the aggregated noisy gradient but cannot deduce any individual client's gradient because of secure aggregation. Intermediate servers receive only one share per client and cannot learn the underlying update. Local clients see only the global model parameters and, under LDP, cannot reconstruct other clients' data.
\end{enumerate}

Based on this analysis, we conclude that DDP-SA protects client data throughout the entire federated learning process and provides end-to-end privacy protection. By combining local perturbation with secure aggregation, DDP-SA reduces the risk of privacy leakage more effectively than either LDP or MPC alone, while maintaining controllable accuracy loss.

\subsubsection{Analysis of Privacy Inference Attacks}
We now discuss several common types of privacy inference attacks in the context of MPC, LDP, and DDP-SA.

\begin{enumerate}[leftmargin=*]
    \item \textbf{Membership inference attacks:} By adding noise to client updates under local differential privacy, DDP-SA and LDP prevent adversaries from reliably determining whether a specific sample was used in training. The noise masks the contribution of individual records, which mitigates membership inference attacks.
    
    \item \textbf{Property inference attacks:} DDP-SA and LDP perturb gradients before aggregation, which hides fine-grained patterns that might reveal sensitive properties of the training data that are not explicitly modeled. This significantly reduces the effectiveness of property inference attacks.
    
    \item \textbf{Training data or label inference attacks:} Secure aggregation in DDP-SA and MPC ensures that the model updates visible to the parameter server are aggregated and not attributable to any single client. This makes it difficult to reconstruct training inputs or labels from observed updates.
    
    \item \textbf{Class representative attacks:} By obfuscating individual gradients through LDP and only revealing aggregates through secure aggregation, DDP-SA and LDP prevent adversaries from reconstructing representative samples for a particular class from the observed gradients.
\end{enumerate}

In summary, DDP-SA is designed to mitigate a wide range of privacy inference attacks, including membership inference, property inference, training data or label inference, and class representative attacks. By combining local differential privacy with secure aggregation, it offers stronger protection than LDP or MPC used in isolation.

\begin{figure*}[!t]
\centering
\subfloat[]{\includegraphics[width=2.95in]{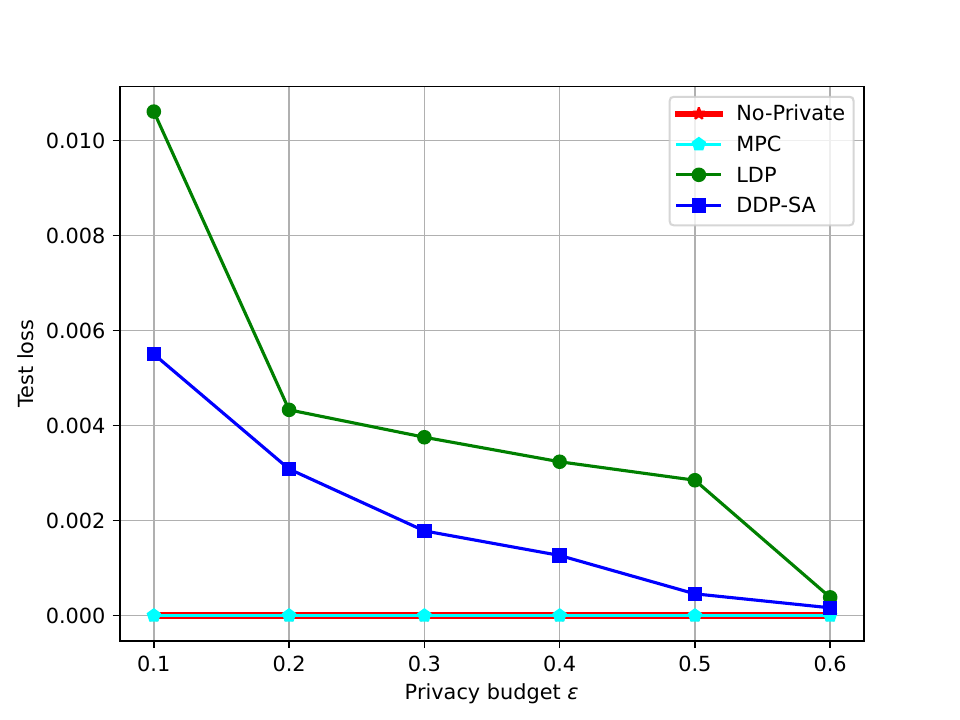}%
}
\hfil
\subfloat[]{\includegraphics[width=2.95in]{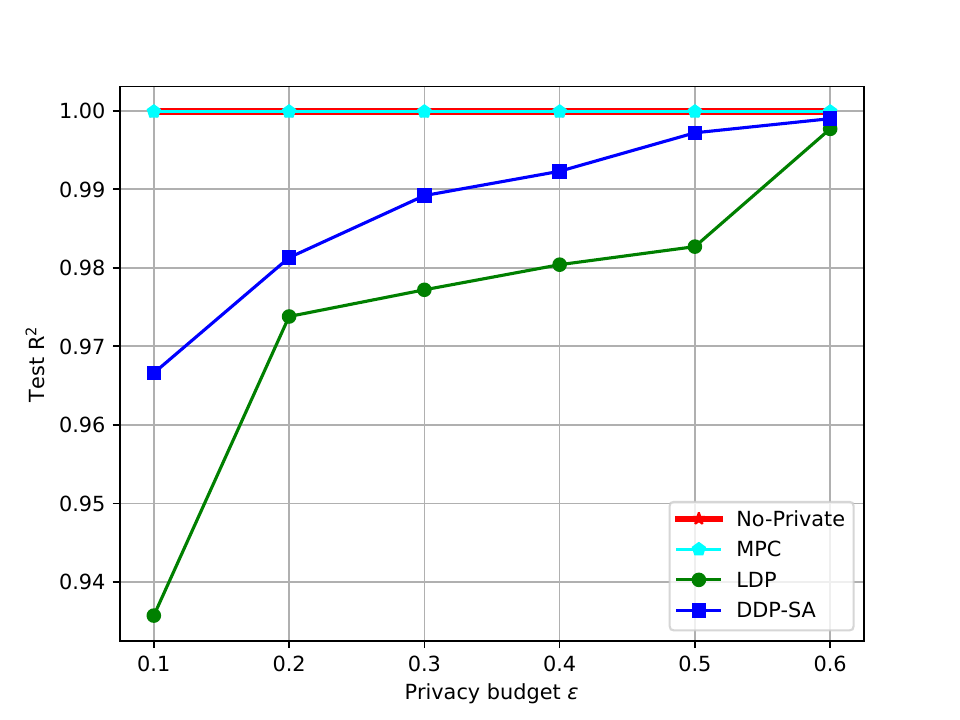}%
}
\caption{Accuracy as a function of privacy budget $\epsilon$. (a) Test loss vs.\ $\epsilon$. (b) Test $\text{R}^2$ vs.\ $\epsilon$.}
\label{Figure 8}
\end{figure*}

\begin{figure*}[!t]
\centering
\subfloat[]{\includegraphics[width=2.95in]{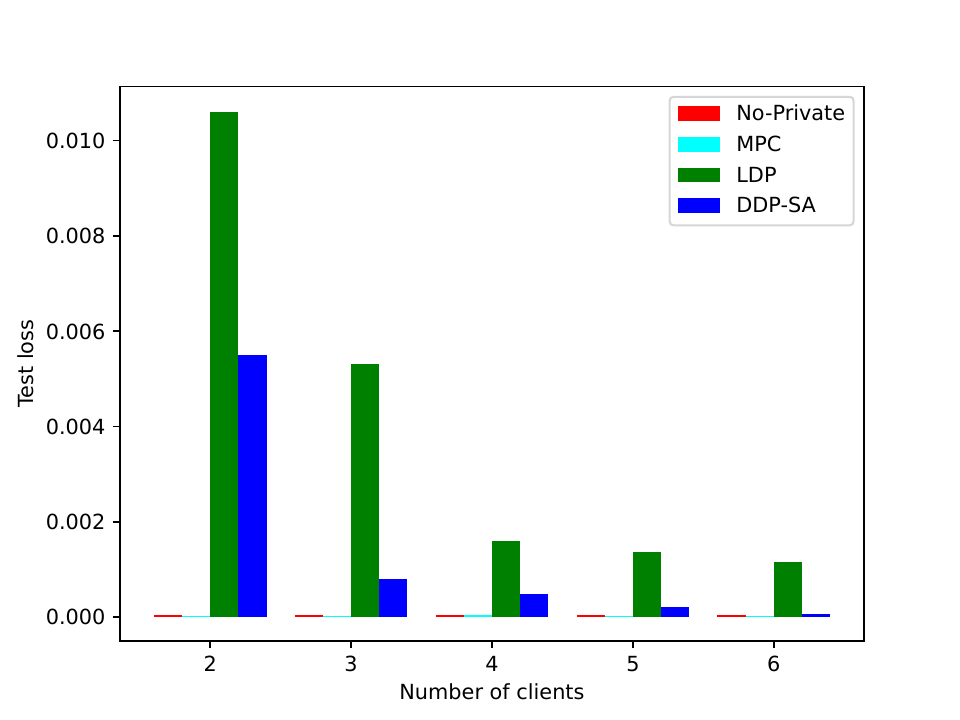}%
}
\hfil
\subfloat[]{\includegraphics[width=2.95in]{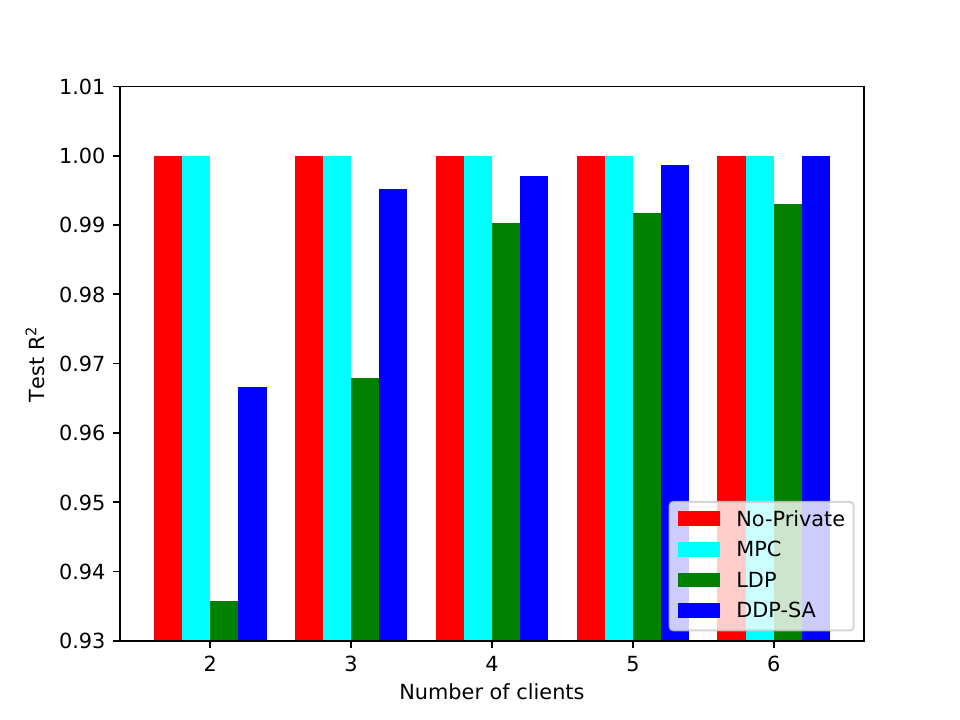}%
}
\caption{Accuracy as a function of the number of clients $n$. (a) Test loss vs.\ $n$. (b) Test $\text{R}^2$ vs.\ $n$.}
\label{Figure 9}
\end{figure*}

\begin{figure}[t]
  \centering
  \includegraphics[width=3.2in, trim=1cm 0.5cm 2cm 0.5cm]{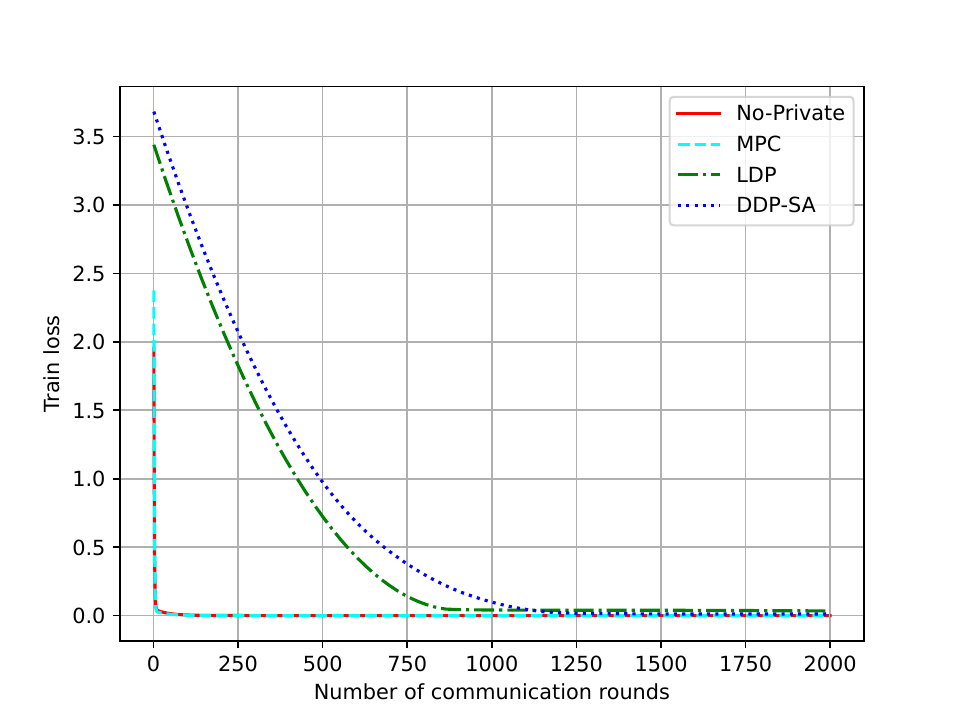}
  \caption{Training loss as a function of the number of communication rounds $T$.}
  \label{Figure 10}
\end{figure}

\subsection{Performance Analysis}
To evaluate the performance of the proposed DDP-SA scheme under varying conditions, we consider three key factors that influence the accuracy of the global model: the privacy budget $\epsilon$, the number of clients $n$, and the number of communication rounds $T$.

\subsubsection{Evaluation with respect to \texorpdfstring{$\epsilon$}{epsilon}}
Fig.~\ref{Figure 8} shows how different values of $\epsilon$ affect model accuracy. In this experiment, $\epsilon$ is varied from 0.1 to 0.6, while all other settings remain fixed. From Fig.~\ref{Figure 8}(a), the test loss of both No-Private and MPC remains close to zero (around $10^{-12}$) for all values of $\epsilon$. The test loss of LDP and DDP-SA decreases as $\epsilon$ increases, and the loss for DDP-SA is consistently lower than that for LDP. When $\epsilon$ reaches 0.6, the test loss of both LDP and DDP-SA is close to $10^{-4}$.

From Fig.~\ref{Figure 8}(b), the test $\text{R}^2$ of No-Private and MPC remains at 0.9999 for all values of $\epsilon$. The test $\text{R}^2$ of LDP and DDP-SA increases with $\epsilon$, and the value for DDP-SA is always higher than that for LDP. When $\epsilon$ reaches 0.6, the test $\text{R}^2$ of both LDP and DDP-SA is close to 0.9999.

These observations reflect the fundamental trade-off in differential privacy. Larger $\epsilon$ implies weaker privacy but higher accuracy, whereas smaller $\epsilon$ implies stronger privacy but lower accuracy. Overall, Fig.~\ref{Figure 8} shows that DDP-SA achieves better accuracy than LDP for all tested values of $\epsilon$.

\subsubsection{Evaluation with respect to \texorpdfstring{$n$}{n}}
The number of participating clients can also affect model accuracy. In this experiment, the number of clients $n$ is increased from 2 to 6 while keeping all other settings fixed. Fig.~\ref{Figure 9} shows the resulting accuracy.

From Fig.~\ref{Figure 9}(a), the test loss of No-Private and MPC remains close to zero (about $10^{-12}$) for all values of $n$. The test loss of LDP and DDP-SA decreases as $n$ increases, and the loss for DDP-SA is always lower than that for LDP. When $n = 6$, the test loss of DDP-SA is close to $10^{-6}$.

Fig.~\ref{Figure 9}(b) shows that the test $\text{R}^2$ of No-Private and MPC remains close to 1 for all values of $n$. The test $\text{R}^2$ of LDP and DDP-SA increases with $n$, and DDP-SA consistently achieves higher $\text{R}^2$ than LDP. When $n = 6$, the test $\text{R}^2$ of DDP-SA is close to 1.

This behavior can be explained by the averaging effect of noise. As the number of clients increases, the average of the added noise tends to zero, and the average noisy gradient approaches the true average gradient. Consequently, the resulting model parameters become closer to the true parameters, which reduces test loss and increases test $\text{R}^2$. Overall, Fig.~\ref{Figure 9} shows that DDP-SA achieves better accuracy than LDP as the number of clients increases.

\subsubsection{Evaluation with respect to \texorpdfstring{$T$}{T}}
Fig.~\ref{Figure 10} shows the effect of the number of communication rounds $T$ on model accuracy. The training loss decreases rapidly as $T$ increases. The number of communication rounds required for convergence is 1041 and 1035 for No-Private and MPC, and 1222 and 1218 for LDP and DDP-SA, respectively. Thus, LDP and DDP-SA require more rounds to reach convergence. Furthermore, the final training loss of DDP-SA is lower than that of LDP.

The increase in required rounds for LDP and DDP-SA is due to the noise added to local gradients, which introduces randomness into the optimization trajectory. This requires more iterations to reach a stable solution. Nevertheless, once converged, DDP-SA achieves better accuracy than LDP, as shown by the lower training loss.

In summary, the performance analysis shows that DDP-SA achieves better accuracy than LDP as the privacy budget $\epsilon$, the number of clients $n$, and the number of communication rounds $T$ increase, while still providing stronger privacy guarantees.

\section{Conclusion}\label{sec:section7}
In this paper, we proposed DDP-SA, a novel privacy-preserving federated learning framework designed to address privacy leakage in the federated learning process. The framework integrates local differential privacy and secure multi-party computation to protect clients' gradients during training, thereby offering stronger defense against privacy inference attacks. Extensive experimental results demonstrate that DDP-SA provides enhanced privacy guarantees compared to using LDP or MPC alone, while maintaining acceptable efficiency and accuracy. In addition, DDP-SA safeguards clients' private data throughout the entire federated learning workflow and effectively mitigates various types of privacy inference attacks.

We also analyzed the performance of DDP-SA under different conditions and showed that it offers superior utility compared to LDP-based approaches. Future work includes exploring optimization strategies to further improve model accuracy and training efficiency, as well as extending the framework to non-IID data distributions and a wider range of model architectures.

\bibliographystyle{IEEEtran}
\bibliography{bibliography}

\end{document}